\documentclass[preprint,11pt,numbers,times]{elsarticle}

\usepackage{amssymb}
\usepackage{array}
\usepackage{multirow}
\usepackage{makecell}
\usepackage{float}
\usepackage{graphicx} 
\usepackage{afterpage}
\usepackage{hyperref}
\usepackage{tabularx}
\usepackage{enumitem}
\usepackage{boldline}
\usepackage{amsmath}
\usepackage{booktabs}
\usepackage{lineno}
\usepackage{nomencl}
\makenomenclature

\renewcommand{\nomgroup}[1]{%
  \ifthenelse{\equal{#1}{A}}{\item[\textbf{Thermal Affordance Assessment}]\vspace{2ex}}{%
  \ifthenelse{\equal{#1}{B}}{\item[\textbf{Thermal Comfort Assessment}]\vspace{2ex}}{%
  \ifthenelse{\equal{#1}{C}}{\item[\textbf{Statistical Modelling}]\vspace{2ex}}{}}}%
}

\usepackage{url}
\usepackage{boxedminipage}
\usepackage{textpos}

\journal{Building and Environment}

\begin{document}

\begin{frontmatter}

\title{Thermal Comfort in Sight: Thermal Affordance and its Visual Assessment for Sustainable Streetscape Design}

\author[inst1,inst5]{Sijie Yang}
\author[inst2]{Adrian Chong}
\author[inst3]{Pengyuan Liu}
\author[inst1,inst4]{Filip Biljecki\corref{cor1}}

\cortext[cor1]{Corresponding author}
\ead{filip@nus.edu.sg}

\affiliation[inst1]{organization={Department of Architecture, National University of Singapore},%
            country={Singapore}}
\affiliation[inst5]{organization={School of Engineering and Applied Science, University of Pennsylvania, Philadelphia},%
            country={United States}}
\affiliation[inst2]{organization={Department of the Built Environment, National University of Singapore},%
            country={Singapore}}

\affiliation[inst3]{organization={Future Cities Lab Global, Singapore-ETH Centre},%
            country={Singapore}}
            
\affiliation[inst4]{organization={Department of Real Estate, National University of Singapore},%
            country={Singapore}}

\begin{abstract}
\begin{textblock*}{\textwidth}(0cm,-12cm)
\begin{center}
\begin{footnotesize}
\begin{boxedminipage}{1\textwidth}
This is the Accepted Manuscript version of an article published by Elsevier in the journal \emph{Building and Environment} in 2025, which is available at:\\ \url{https://doi.org/10.1016/j.buildenv.2025.112569}\\ Cite as:
Yang S, Chong A, Liu P, Biljecki F (2025): Thermal comfort in sight: Thermal affordance and its visual assessment for sustainable streetscape design. \textit{Building and Environment}, 271: 112569.
\end{boxedminipage}
\end{footnotesize}
\end{center}
\end{textblock*}

\begin{textblock*}{1.5\textwidth}(-0.8cm,15cm)
{\tiny{\copyright{ }2025, Elsevier. Licensed under the Creative Commons Attribution-NonCommercial-NoDerivatives 4.0 International (\url{http://creativecommons.org/licenses/by-nc-nd/4.0/})}}
\end{textblock*}
In response to climate change and urban heat island effects, enhancing human thermal comfort in cities is crucial for sustainable urban development. Traditional methods for investigating the urban thermal environment and corresponding human thermal comfort level are often resource intensive, inefficient, and limited in scope. To address these challenges, we (1) introduce a new concept named thermal affordance, which formalizes the integrated inherent capacity of a streetscape to influence human thermal comfort based on its visual and physical features; and (2) an efficient method to evaluate it (visual assessment of thermal affordance --- VATA), which combines street view imagery (SVI), online and in-field surveys, and statistical learning algorithms. VATA extracts five categories of image features from SVI data and establishes 19 visual-perceptual indicators for streetscape visual assessment. Using a multi-task neural network and elastic net regression, we model their chained relationship to predict and comprehend thermal affordance for Singapore. VATA predictions are validated with field-investigated OTC data, providing a cost-effective, scalable, and transferable method to assess the thermal comfort potential of urban streetscape. Moreover, we demonstrate its utility by generating a geospatially explicit mapping of thermal affordance, outlining a model update workflow for long-term urban-scale analysis, and implementing a two-stage prediction and inference approach (IF-VPI-VATA) to guide future streetscape improvements. This framework can inform streetscape design to support sustainable, liveable, and resilient urban environments.

\end{abstract}


\end{frontmatter}

\section{Introduction}
\label{sec:sample1}
\subsection{Urban heat environment and outdoor thermal comfort (OTC)}

\nomenclature[B]{OTC}{outdoor thermal comfort}
\nomenclature[B]{UHI}{urban heat island}
\nomenclature[B]{ITC}{indoor thermal comfort}

The growing effect of urban heat island (UHI) has brought the thermal environment of cities and their adaptive strategies into the spotlight~\citep{alcoforado2008global, li2018urban, takebayashi2015improvement, lai2019review, tseliou2022evaluating}. UHI is detrimental to the built environment -- it drives extreme climate changes \citep{chen2022estimating, steensen2022future}, increases energy consumption for cooling \citep{errebai2022impact}, and deteriorates health due to poor air quality and decreased thermal comfort \citep{tan2010urban}.

Outdoor thermal comfort (OTC), the subjective evaluation of one's satisfaction with urban thermal conditions, is a key area of study \citep{chen2012outdoor, shooshtarian2020outdoor}. Poor OTC that makes people feel hotter can raise skin temperatures, increasing the risk of heat-related illness \citep{da2023assessment, kovats2008heat}, and can even affect mental health \citep{piracha2022urban, hwong2022climate}. The growing concern around OTC has driven the need for effective OTC evaluation methods \citep{aghamolaei2022comprehensive}. However, traditional methods, which focus on field surveys and environmental measurements, remain costly and limited in scalability and precision \citep{johansson2014instruments, evola2020novel, xiong2022spatiotemporal}. The research by \cite{li2024sensitivity} offers a promising example of high-resolution thermal climate mapping by integrating urban morphology and meteorological data. Nevertheless, it lacks detailed consideration of the street environment and does not incorporate survey or measurement validation of human comfort at the street level. These constraints, amid climate change’s growing impact, call for improved methods to assess urban-scale OTC, for fostering sustainable street designs that enhance thermal comfort and urban livability. 

Urban environments, characterised by their intricate and complex nature, deviate significantly from the finite conditions of indoor thermal comfort (ITC) evaluation. The multitude of diverse variables present in urban environments, such as solar radiation and shade availability, complicates climatic parameter control in OTC evaluations \citep{spagnolo2003field}. This complexity hinders comprehensive OTC assessments in urban areas \citep{honjo2009thermal}. To address this challenge, \cite{mayer1987thermal} developed the physiological equivalent temperature (PET) for estimating outdoor thermal conditions comparable to indoor environments. Another model applied for OTC is the predicted mean vote (PMV), developed by \cite{fanger1970thermal}, which predicts the average thermal sensation on a scale from -3 (cold) to +3 (hot), based on six environmental factors: air temperature, radiant temperature, relative humidity, air speed, clothing insulation, and metabolic rate. Recently, the universal thermal climate index (UTCI) has gained attention for its dynamic consideration of human physiological responses, making it particularly suitable for urban environments \citep{jendritzky2012utci, brode2012predicting}. PET, PMV, and UTCI are now widely used for OTC evaluation, though the complexity of urban meteorological variables and human factors makes the process challenging \citep{honjo2009thermal}. 

Traditional OTC measurements based on these indexes rely on meteorological instruments, field investigations, and surveys \citep{acero2015comparison, mirzaei2015recent}, but these methods are resource-intensive, limited in spatial coverage, and lack geographical precision \citep{aghamolaei2022comprehensive}. Such limitations highlight the need for a scalable, cost-effective framework for urban-scale OTC assessment that integrates multidimensional data while ensuring precision and applicability in sustainable urban planning. Moreover, recent advances in ITC research have introduced active learning approaches to thermal comfort modelling, enabling more adaptive and data-efficient solutions \citep{abdelrahman2022personal, tekler2023hybrid, ono2022effects}. These developments, coupled with the growing demand for responsive planning frameworks, have paved the way for enhanced thermal comfort assessment systems that not only integrate changing environmental factors but also align with dynamic urban planning needs \citep{silva2018urban, batty2024digital}. Together, these trends underline the shift toward a more efficient, responsive and adaptive framework for urban-scale OTC evaluation, which is essential for sustainable streetscape design and human-centric urban development in the digital era.

\subsection{Visual assessment of streetscape quality based on street view imagery (SVI)}

\nomenclature[A]{SVI(s)}{street view imagery}

Street view imagery (SVI) is a crucial tool for assessing the quality of streetscapes and their social impacts in urban environments \citep{gilge2016google,biljecki2021street, qiu2022subjective, liang2023revealing, liang2024evaluating, zhao2024exploring}. The visual composition of streetscapes provides a wealth of multidimensional information, enabling researchers to understand the socioeconomic implications of built environments and guide improvements in urban design \citep{lee2023current}. SVI, paired with computer vision and GIS, has become a standard method for large-scale streetscape analysis \citep{cheng2017use, kang2020review, larkin2021predicting, ki2023bridging, hou2024sensing, fan2024nighttime}. It supports both objective analysis of streetscape elements, including \textbf{image features (IF)} such as geometry, colour, composition, and texture, and captures subjective \textbf{visual-perceptual indicators (VPI)} such as visual comfort, enclosure, complexity, through online surveys \citep{naik2014streetscore, xu2022associations, zhao2023sensing}. IF refers to the objective, measurable aspects of an image (SVI in urban studies) that can be extracted and analysed through computational techniques \citep{xiao2009multiple, nagata2020objective, yap2023incorporating, wei2023integrating, fujiwara2024microclimate}, while VPI, on the other hand, refers to the subjective, human-centred interpretations of streetscapes, capturing how people perceive and experience the urban environment \citep{kang2020review, ogawa2024evaluating, ito2024understanding}. 

SVI captures a variety of IFs that can be extracted using advanced computer vision techniques. Algorithms like PSPNet \citep{zhao2017pyramid} and FCN \citep{long2015fully}, trained on datasets such as Cityscapes \citep{cordts2016cityscapes} and ADE20K \citep{zhou2017scene}, enable precise semantic segmentation of SVI. These algorithms allow researchers to quantify the proportions of elements such as buildings, trees, and roads, and open-source packages are developed for semantic SVI data analytics in urban planning \citep{yap2022free, yap2023urbanity, ito2024zensvi}. In addition to streetscape indices based on element percentages, other metrics have been developed to capture additional IFs of urban streetscapes in SVI. These include vehicle and pedestrian counts \citep{goel2018estimating}, the architectural style and building characteristics \citep{kang2018building, xu2023urban, lei2024predicting}, urban scene types \citep{su2021urban}, and weather conditions \citep{ibrahim2019weathernet}. These metrics, derived from specialised computer models and datasets, provide valuable spatial data for assessing urban environment.

Beyond objective IFs derived from computer vision and SVI datasets, human visual assessments of streetscapes involve subjective VPIs, which are often interrelated \citep{qiu2022subjective, yang2023role}. These subjective qualities can be evaluated using SVI-based VPI questionnaires \citep{ogawa2024evaluating}, typically conducted online. Participants rate or compare SVIs based on classic subjective VPIs such as imageability, enclosure, human scale, transparency, complexity, and safety, reflecting their perceptions of urban spaces \citep{ewing2009measuring, qiu2022subjective, xu2022associations}. Many studies have invited participants to perform pairwise comparisons or grading-based (such as score range 1-10) visual assessments of SVI to evaluate streetscape quality and subjective visual perception they evoke \citep{kang2020review, larkin2021predicting, ito2024translating}.

As the urban thermal environment and corresponding thermal comfort are closely linked to the objective composition of streetscapes and subjective visual perceptions \citep{nikolopoulou2011outdoor, klemm2015street, yilmaz2021street, hu2024street, yu2024combining}, SVI has proven to be an important instrument in this domain. For instance, \cite{hu2024street} uses 3D morphological data from SVI for street-level thermal mapping, while \cite{yu2024combining} combines SVI features with urban social-physical data for seasonal thermal assessments. When it comes to using streetscape elements and visual perception for thermal comfort evaluation, \cite{klemm2015street} identifies three key ways street greenery enhances thermal comfort across different street types, based on SVI and micrometeorological measurements: lowering radiant temperatures, improving streetscape aesthetics, and positively influencing conscious perceptions of microclimate. Similarly, after the evaluation of streetscape design through satellite images and SVI, local climate conditions, and ENVI-met modelling data, \citep{yilmaz2021street} suggests that thermal comfort can be improved through open spaces and thoughtful planting choices. Therefore, comprehensive visual assessments of streetscape quality not only enhance aesthetics and functionality but also improve thermal comfort and urban liveability.

\subsection{Affordance information from SVI in urban studies}
The theory of affordance, proposed by \cite{gibson1977theory}, highlights how the ecological environment shapes perception and behaviour. Applied to the built environment, affordances represent environmental features that promote human perception and behaviour \citep{gibson1977theory, shashank2022creation}. According to \cite{gibson1977theory}, the environment acts as a ‘surface’ medium through which humans perceive and interact with their surroundings, offering clues to potential actions. The informational content and environmental characteristics embedded within the 'surfaces' constitute the ‘value’ and ‘meaning’ of a place. These features, defined as affordances and inferred from environmental attributes, indicate likely human perceptions and behaviours even without direct experience. \cite{gibson1977theory} further emphasises that this inherent affordance of the environment can be deduced through visual assessment.

Affordance theory has been employed to elucidate the relationship between the built environment and human mobility behaviour, utilising geospatial indicators such as greenery index, street network morphology metrics, and urban function density, to comprehend how environments support activities like walking, biking, and running \citep{fusco2016beyond, shashank2022creation, ito2024translating, yang2024effect}. The built environment provides a foundation for urban life, influencing social activities, events, human movement, and exploration. Understanding these affordances is crucial for improving city conditions. SVI, as a direct and human-centred data source, offers valuable insights into the built environment’s affordances. Until now, in conjunction with SVI, affordance theory has been used to explain the evaluations of playability \citep{kruse2021places}, runnability \citep{shashank2022creation}, and spatial navigation \citep{gregorians2022affordances}, which has been shown to be a robust theoretical framework for explaining the relationship between urban attributes and human perception or behaviour \citep{shashank2022creation}.

Both objective SVI indices and its subjective VPIs derived from human visual assessment offer insights into the affordances of urban streetscapes. For example, greenery indices derived from these images link higher green views to increased pedestrian activity and improved mental well-being \citep{wang2019urban, he2020association}, while visual complexity affects perceived safety and comfort \citep{kawshalya2022impact}. By combining objective SVI analysis with subjective SVI assessment surveys, we can explore various urban spaces and their perceptual and behavioural potentials across different urban scales \citep{wang2023drivers, yang2023role}. This affordance analysis framework, using both objective and subjective metrics, effectively evaluates urban environments and streetscape design \citep{song2022coherence, qiu2023subjective}. These assessments can reveal a variety of human behaviours and urban outcomes, including walking and cycling patterns \citep{ito2021assessing}, property values \citep{qiu2023subjective, yang2023role}, and public health indicators \citep{larkin2019evaluating, wang2019relationship}. Superior streetscape design improves the visual experience, walkability, bikeability, environmental appeal, and public activities, boosting the value of urban streets \citep{park2014perception, gan2021associations, strath2012measured}.

\subsection{Research gaps and our response}
The interrelation between thermal comfort and streetscape design constitutes a significant area of study \citep{jamei2016review}. Based on existing literature \citep{klemm2015street, kang2020review, yilmaz2021street, hu2024street, yu2024combining}, we posit that both the objective composition of streetscape elements and subjective visual assessment could serve as proxies for evaluating a streetscape’s capacity to influence human thermal comfort. However, key research gaps persist.

Firstly, there is no clear concept for assessing the built environment’s objective properties that influence thermal comfort in streetscape design. Although many studies have explored associations between streetscape elements, outdoor thermal environment, and OTC \citep{labdaoui2021street, sadeghi2021urban, zheng2024evaluation, yu2024combining, hu2024street}, a theoretical framework that explains these relationships remains underdeveloped. This limits the creation of a convincing framework for sustainable streetscape design that improves outdoor thermal comfort.

To address this gap, we introduce the concept of \textbf{thermal affordance} (see details in Section 2), inspired by Gibson's theory of environmental affordance \citep{gibson1977theory}. Our novel concept highlights the potential of urban design to promote thermal comfort through streetscape element configurations, forming the foundation for a data-driven framework to analyse their impact on OTC.

Secondly, the use of human visual assessment in evaluating streetscapes' thermal conditions and their capability to impact thermal comfort is rarely studied. While some research has touched on the interaction between visual information from SVI and outdoor thermal environments \citep{klemm2015street, urban2022using, zhang2022effects, yan2023mediating, hu2024street}, they have not applied human visual assessment to examine how streetscape element composition influences thermal comfort. Some studies predict thermal environments using SVI \citep{yu2024combining, hu2024street}, but they overlook OTC and lack validation through field experiments. Therefore, a methodological approach incorporating SVI and human visual assessment is needed for urban-scale OTC evaluation, validated by in-field OTC measurements. Moreover, while it is widely accepted that street greenery improves OTC \citep{klemm2015street, jamei2016review, motie2023assessment}, the influence of various streetscape elements on perceived thermal comfort remains underexplored, especially regarding buildings, roads, and vehicles.

Our proposed \textbf{visual assessment of thermal affordance (VATA)} framework fills this gap by using SVI data and online questionnaires to quantify thermal affordance by integrating \textbf{image features (IF)} and human \textbf{visual-perceptual indicators (VPI)} from online surveys. VATA provides a novel approach to urban OTC evaluation.

Third, the systematic application of multi-source urban data, such as SVI, human visual-perceptual surveys, and in-field comfort and physiological indicators, has not been fully adopted as an efficient workflow for assessing OTC in urban environments. Although some studies have linked SVI features with the urban thermal environment \citep{urban2022using, yan2022influence, kim2023heat, hu2024street, klimenko2024instant}, a verifiable, replicable, and street-level methodology for urban-scale OTC evaluation supported by multi-source urban data has not yet been built. 

To address this methodological shortfall, our VATA framework introduces a data-driven workflow that integrates SVI, human visual assessments, and in-field comfort data. This approach enables the development and validation of prediction and inference models based on our proposed concept of thermal affordance, facilitating scalable and precise OTC evaluation for sustainable urban planning.

\subsection{Research question}
Given the identified research gaps, we propose the following research questions:

\begin{enumerate}
    \item How can the concept of thermal affordance be utilised as a theoretical framework to describe and assess urban street environment properties that promote relative thermal comfort potential?
    \item How can the visual assessment of thermal affordance (VATA) framework be constructed based on elements that can be extracted from street view images (SVI) --- image features (IF) and human visual-perceptual indicators (VPI) --- to perform prediction and inference models to efficiently assess and understand urban-scale streetscapes’ OTC potential (thermal affordance) for sustainable urban planning?
    \item How can we develop a multi-source data analysis workflow, incorporating in-field outdoor thermal comfort (OTC) investigation data, to validate the VATA framework?
    \item What analytical insights and urban planning frameworks can be derived from the data outputs of the VATA framework to guide sustainable streetscape design?
\end{enumerate}

\section{The idea of thermal affordance}

\subsection{Proposed definition of thermal affordance}

    \begin{figure}[!ht]
        \centering 
        \includegraphics[width=1\textwidth]{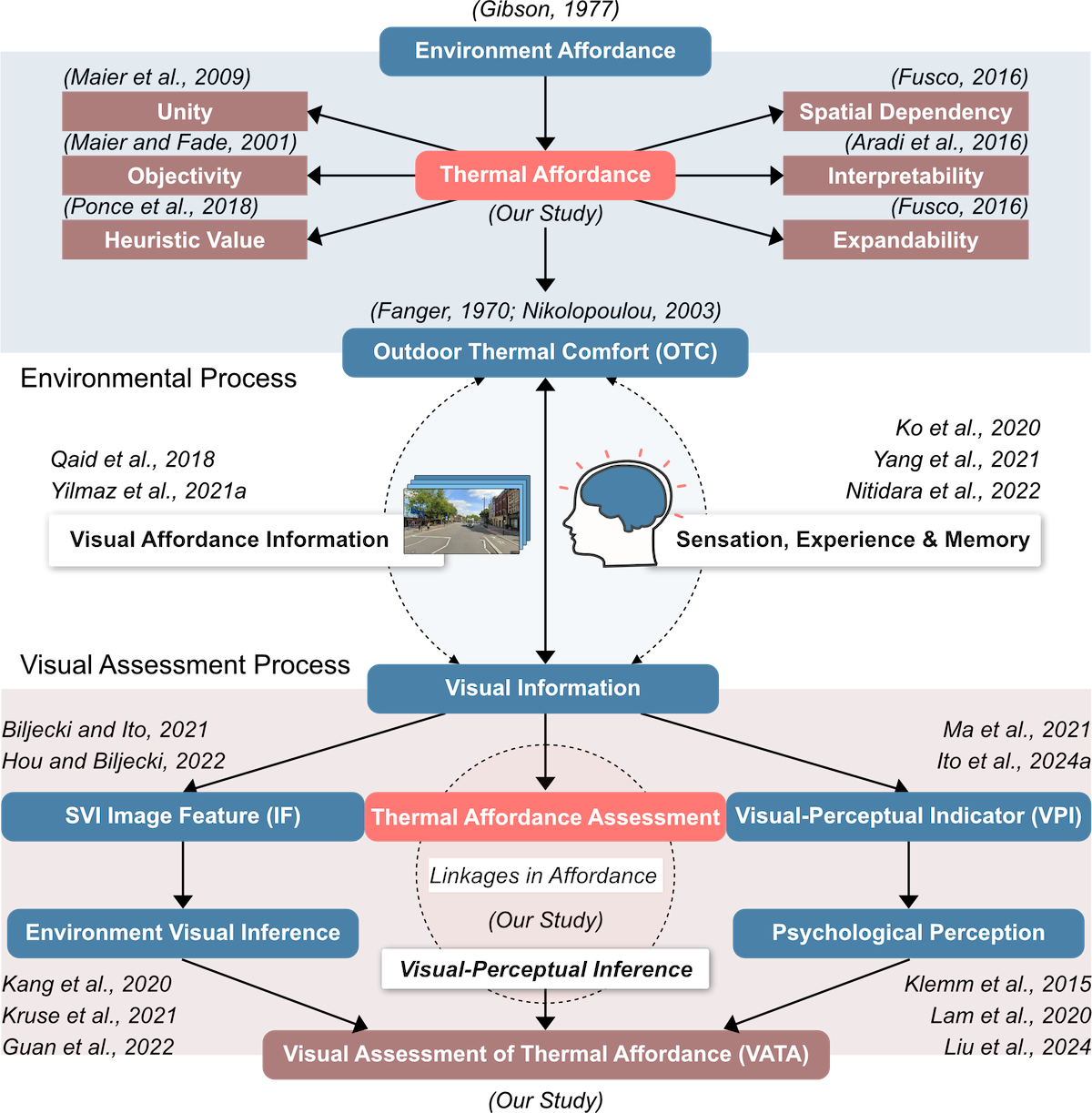} 
        \caption{VATA conceptual framework: from thermal affordance to its visual assessment. Source of the SVI: Google Street View.}
        \label{VATA_concept}
    \end{figure}

As Gibson’s theory of affordance suggests \citep{gibson1977theory}, environments contain inherent values and information that shape human perceptions and behaviours. Thermal comfort, as part of human perceptions, reflects both subjective satisfaction and objective factors such as air temperature, air humidity, and wind speed \citep{fanger1970thermal, nikolopoulou2003thermal, van2010thermal, nicol2017rethinking, de2021housing}. We introduce the concept of \textbf{'thermal affordance'} to describe the inherent capability of an environment to impact thermal comfort. This concept integrates various environmental factors, indicating the possible thermal comfort experienced. Studies have shown the connection between thermal comfort and integrated environmental factors such as walkability \citep{labdaoui2021street} and streetscape design \citep{yilmaz2021street}, supporting the validity of thermal affordance.

Currently, the literature on thermal comfort presents various interconnected concepts, each providing a different lens to understand how individuals experience thermal environments \citep{lai2020comprehensive}. For instance, thermal comfort is defined by the ASHRAE Standard 55 as a 'condition of mind that expresses satisfaction with the thermal environment', emphasising a subjective evaluation \citep{standard2023thermal}. Meanwhile, thermal neutrality refers to the conditions where the majority of individuals feel neither warm nor cold \citep{shooshtarian2017study, standard2023thermal}. Concepts such as the thermoneutral zone and thermal sensation further highlight the physiological and psychological dimensions of thermal comfort \citep{bligh1973glossary}, with the latter focussing on how people consciously perceive temperature \citep{parsons2007human, standard2023thermal}.

These established concepts offer significant insights, but they remain largely subjective, relying heavily on individuals’ preferences or perceptions. In contrast, our proposed notion of thermal affordance aims to transcend individual subjectivity by focussing on the inherent characteristics of the environment itself and its potential to influence thermal comfort. While more concepts like thermal expectation or thermal satisfaction consider personal inclinations and thermal experiences \citep{de2001adaptive, nikolopoulou2003thermal, dzyuban2022outdoor}, thermal affordance unifies objective environmental features, offering a more scalable and integrative framework for OTC improvement in urban planning and design.

The necessity of introducing thermal affordance lies in its ability to bridge the gap between subjective experiences of thermal environments and the objective environmental factors that shape them. Such an approach allows for a more comprehensive analysis, helping urban planners make informed decisions that improve thermal comfort by connecting measurable environmental properties with subjective or individual-based assessments.
    
We summarise several key characteristics of thermal affordance: unity, objectivity, heuristic value, spatial dependency, interpretability, and expandability, as illustrated in Figure~\ref{VATA_concept}.
    
    \begin{itemize}
        \item \textbf{Unity} indicates that thermal affordance aims to encompass all fixed variables that influence thermal comfort in the environment into a unified set for description, based on the integration and unity properties of affordance proposed by \cite{maier2009affordance}.
        \item \textbf{Objectivity} highlights the objective environmental information encapsulated in thermal affordance, unaffected by individuals' subjective perspectives, aligning with the idea of affordance as an integration of objectivity in the environment \citep{maier2001affordance}.
        \item \textbf{Heuristic value} of thermal affordance suggests that the information it encompasses can inspire understanding and analysis of thermal comfort, and the various dimensions of information within thermal affordance can mutually infer and enlighten one another. For instance, \cite{ponce2018deep} suggests the possibility that visual data serve as heuristics for thermostats.
        \item \textbf{Spatial dependency} highlights variations in thermal comfort caused by different spatial environments. Differences in spatial or environmental properties have been shown to directly lead to a difference in affordance \citep{fusco2016beyond}. Thermal affordance does not stress absolute levels of thermal comfort, but rather emphasises spatial differences between environments in their capacity to promote thermal comfort.
        \item \textbf{Interpretability} means that thermal affordance has a clear inference relationship with environmental attributes, as environmental affordance is composed of multiple influencing parameters \citep{aradi2016urban}. Comprehending this inference relationship can aid urban planners in enhancing thermal comfort through urban design.
        \item \textbf{Expandability} means that thermal affordance can continuously incorporate additional environmental variables to update information on an environment's capacity to influence thermal comfort. More dimensions of information, such as lifestyle and dwelling mode \citep{fusco2016beyond}, have been shown to allow for a more comprehensive evaluation and analysis of environmental affordance.
    \end{itemize}

\subsection{Visual assessment of thermal affordance (VATA) framework}

\nomenclature[A]{VATA}{visual assessment of thermal affordance}
\nomenclature[A]{IF(s)}{image feature}
\nomenclature[A]{VPI(s)}{visual-perceptual indicator}

    \begin{figure}[!ht]
        \centering 
        \includegraphics[width=1\textwidth]{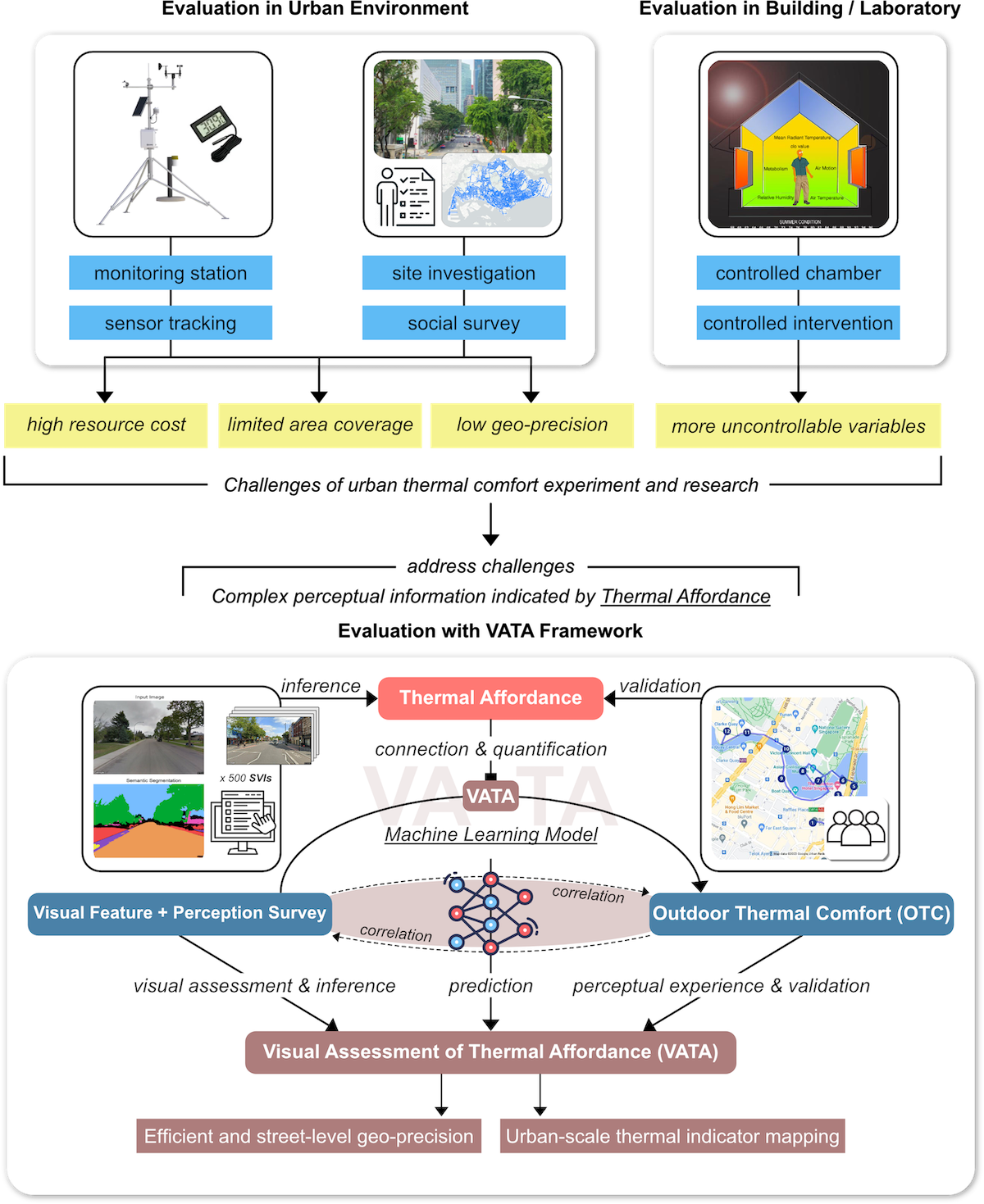} 
        \caption{Visual Assessment of Thermal Affordance (VATA) framework: from challenges to a solution.}
        \label{VATA_framework}
    \end{figure} 

This study identifies four key challenges in urban-scale OTC evaluation (Figure~\ref{VATA_framework}): (1) Measurement of thermal comfort requires substantial resources -- labour, budget, and time \citep{kumar2020study}. (2) Current methods, relying on physical instruments and surveys, can only cover limited urban areas \citep{balogun2019outdoor}. (3) Geographic precision is constrained for city-wide OTC assessments \citep{mutani2022geospatial}. (4) Urban OTC measurements are affected by many uncontrollable environmental variables, unlike controlled indoor experiments \citep{chen2012outdoor}. To address these challenges, we developed the \textbf{VATA (visual assessment of thermal affordance)} framework to enhance urban-scale OTC assessments and improve understanding of urban thermal environments.

This research continues to explore the intrinsic connection between OTC in urban environments and visual data from SVI, shaped by affordance information and personal experiences. First, SVI visual data reflect objective characteristics of streetscapes, indicating streetscapes' potential capability to promote thermal comfort and microclimate conditions \citep{qaid2018effect, yilmaz2021analysis}, which forms thermal affordance. Second, this visual information is linked to individuals’ thermal perception, influenced by memory and sensory experiences \citep{ko2020impact, yang2021investigation, nitidara2022multisensory}, which affects their thermal affordance assessment. These insights form the foundation of our VATA analysis, highlighting the role of visual characteristics and perceptual experiences in OTC assessment. Using machine learning, we combine SVI \textbf{image features (IF)} and \textbf{visual-perceptual indicator (VPI)} survey data to predict the VATA metric, allowing us to visually assess the urban environments' capability to promote thermal comfort. Our model shows promise for streetscape assessment and VATA prediction, as illustrated in Figure~\ref{VATA_framework}.

Objective IF from SVI reflect the urban environment from an individual’s perspective in a three-dimensional, objective manner \citep{biljecki2021street, hou2022comprehensive}, capturing elements such as greening rate \citep{li2015assessing}, tree canopies\cite{fujiwara2024panorama}, building information \citep{kang2018building}, urban function \citep{zhang2023knowledge}, and urban form \citep{biljecki2023sensitivity}. These features, closely related to OTC, serve as affordance information since OTC is influenced by surrounding urban elements \citep{yang2011thermal, deng2023influence}, enabling visual-environmental inference in urban studies \citep{kang2020review, kruse2021places, guan2022modelling}. The VPI information obtained through the SVI visual assessment surveys further accounts for urban perceptions of individuals \citep{ma2021visualizing, ito2024understanding}, impacting socioeconomic dimensions such as bikeability \citep{ito2021assessing}, running ability \citep{dong2023assessing} and property prices \citep{yang2023role}. VPI and OTC, as psychological perceptions, subtly influence each other \citep{klemm2015street, shooshtarian2017effect, lam2020cross, liu2024psychological}. Figure~\ref{VATA_concept} illustrates the relationships between IF, VPI, VATA and OTC, demonstrating that the use of SVI for the evaluation of VATA and OTC is grounded in both environmental and perceptual principles. These relationships are not directly derived or scored from the survey data, as the problem remains a “black box” with only separate scores available for IF, VPI, VATA, and OTC. Accordingly, machine learning techniques—including both predictive and inferential modelling—are used to uncover and quantify these underlying relationships.

From another perspective, the intricate relationship between VATA and OTC can be deeply rooted in individuals’ long-term memory and experiential knowledge, as people's sensory experience in urban environments is a long-term process. This kind of environmental experience goes beyond short-term observations and involves complex reasoning and memory processes in the brain \citep{knauff2009neuro, ougmen2016new, wang2021indoor}. For example, the association of greenery with shading and cooling stems from prolonged human experiences \citep{bar2009proactive, eichenbaum2009neurobiology}. This knowledge enables individuals to infer OTC from visual cues, similar to how greenery is linked to cooler temperatures and better thermal comfort\citep{friston2003learning, berkman2013beyond, ma2024effects}. Thus, subjective streetscape visual assessments in our VATA framework are also grounded in the brain’s ability to interpret environmental cues based on past experiences.

While VATA provides an efficient means of quantifying thermal affordance using SVI data, it is important to acknowledge that integrating meteorological information, detailed geospatial data, remote sensing imagery, and socio-cultural contexts can yield more comprehensive insights into how multiple factors collectively influence urban thermal comfort \citep{acero2015comparison, sharma2021assessing, wu2023observed}. Incorporating additional environmental variables could further enhance the VATA framework by employing advanced data fusion and machine learning techniques, ultimately improving predictive accuracy and informing more effective urban design strategies.

\section{Methodology}
\subsection{Research framework and study area}

\nomenclature[C]{MTNNL}{multi-task neural network learning}
\nomenclature[C]{ENRM}{elastic net regression model}

    \begin{figure}[!ht]
        \centering    
        \includegraphics[width=1\textwidth]{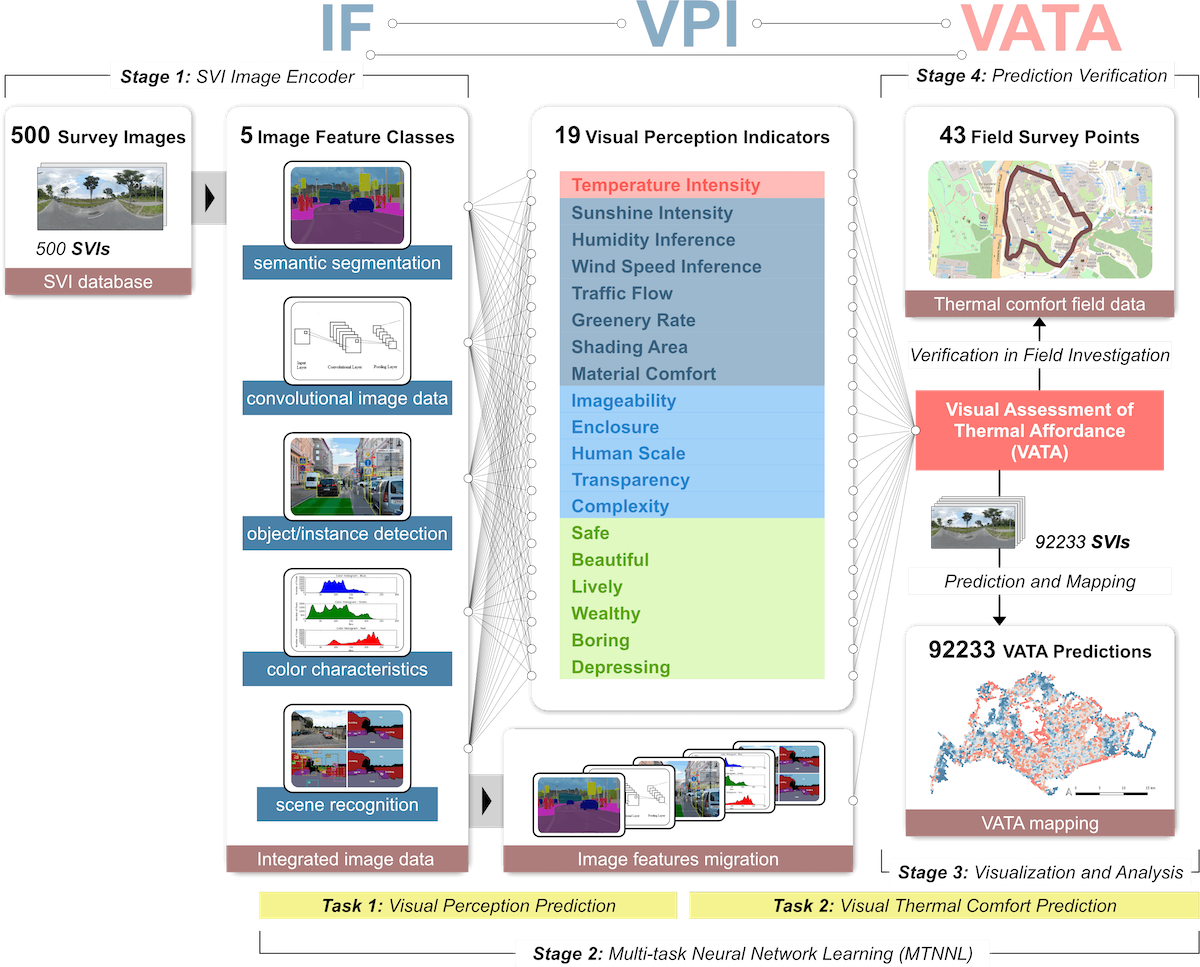} 
        \caption{Research framework based on IFs from SVI data, VPIs from online SVI visual assessment survey, and a two-stage MTNNL model for VATA prediction.} 
        \label{method_framework} 
    \end{figure}

    Figure~\ref{method_framework} presents the research framework for this study, built on the VATA framework. We conducted an online SVI visual assessment survey, evaluating VATA and 19 other VPIs based on 500 SVIs. Five classes of IFs from SVI data, along with survey-based VATA and VPI data, were used to develop datasets for statistical VATA prediction and inference models. Using machine learning models, we can predict and infer relationships between IFs, VPIs, and VATA based on their respective scoring data, all drawn from the same set of 500 SVIs used in the visual assessment survey. A multi-task neural network learning model (MTNNL) was constructed with two stages: predicting VPIs from IFs, and then predicting VATA from VPIs, using weighted loss values for iterative training. This model was applied to predict VATA for 92,233 SVIs in Singapore, and validated against real-world OTC data. A two-stage elastic net regression model (ENRM) was also used to interpret IF-VPI-VATA relationships, offering insights for streetscape design. SVIs are sourced from Google Street View \citep{anguelov2010google}.
    
    Singapore serves as the study area due to its rich urban landscape design, diverse urban forms, consistent outdoor temperatures (25-31 °C year-round) and perennial need to ensure OTC as a tropical city \citep{yang2013thermal}.
    Our study includes an online SVI visual assessment survey, model prediction and inference, and validation using OTC field data, aiming to propose a more efficient method for urban-scale OTC assessment. %

\subsection{Online SVI assessment survey setting}
\label{sec:method_survey}

    \begin{table}[!ht]
    \caption{Selected VPIs and associated questions for the online SVI assessment survey.}
    \label{method_question}
    \centering
    \footnotesize
    \begin{tabularx}{\textwidth}{p{0.45\textwidth} p{0.5\textwidth}}
        \toprule
        \textbf{Visual-Perceptual Indicator (VPI)} & \textbf{Question} \\
        \midrule
        \textbf{\textit{Thermal Affordance Assessment}} & \multirow{2}{=}{Which street view image do you perceive as having a more comfortable outdoor thermal environment for you?} \\
        Visual Assessment of Thermal Affordance (VATA) & \\
        \midrule
        \textbf{\textit{Microclimate Inference}} & \multirow{2}{=}{Which street view image do you perceive exhibits a higher outdoor temperature / sunlight intensity / humidity inference / wind speed inference?} \\
        Temperature Intensity / Sunshine Intensity / Humidity Inference / Wind Speed Inference & \\
        \midrule
        \textbf{\textit{Environmental Evaluation}} & \multirow{2}{=}{Which street view image do you think showcases higher (more) traffic flow / greenery rate / shading areas / construction material comfort?} \\
        Traffic Flow / Greenery Rate / Shading Area / Construction Material Comfort & \\
        \midrule
        \textbf{\textit{Design Quality}} & \multirow{3}{=}{Which street view image stands out to you as a more impressive place / enclosed space / accommodating for human scale / transparent space / complex environment?} \\
        Imageability / Enclosure / Human-Scale / Transparency / Complexity & \\
         & \\
        \midrule
        \textbf{\textit{Evoked Emotion}} & \multirow{2}{=}{Which street view image do you feel evokes a more safe / beautiful / lively / wealthy / boring / depressing atmosphere?} \\
        Safe / Beautiful / Lively / Wealthy / Boring / Depressing & \\
        \bottomrule
    \end{tabularx}
    \end{table}
    
    The study conducts an online SVI visual assessment survey on VATA and 19 other VPIs among 176 long-term Singapore residents using 500 SVIs. The survey aimed to capture residents' subjective visual assessment of thermal affordance-related inferences and perceptions toward various streetscapes shown in random SVIs. These results, combined with IFs, were used to train the MTNNL model to predict the VATA metric, reflecting street-level thermal affordance evaluations.

\begin{figure}[!ht]
        \centering    
        \includegraphics[width=1\textwidth]{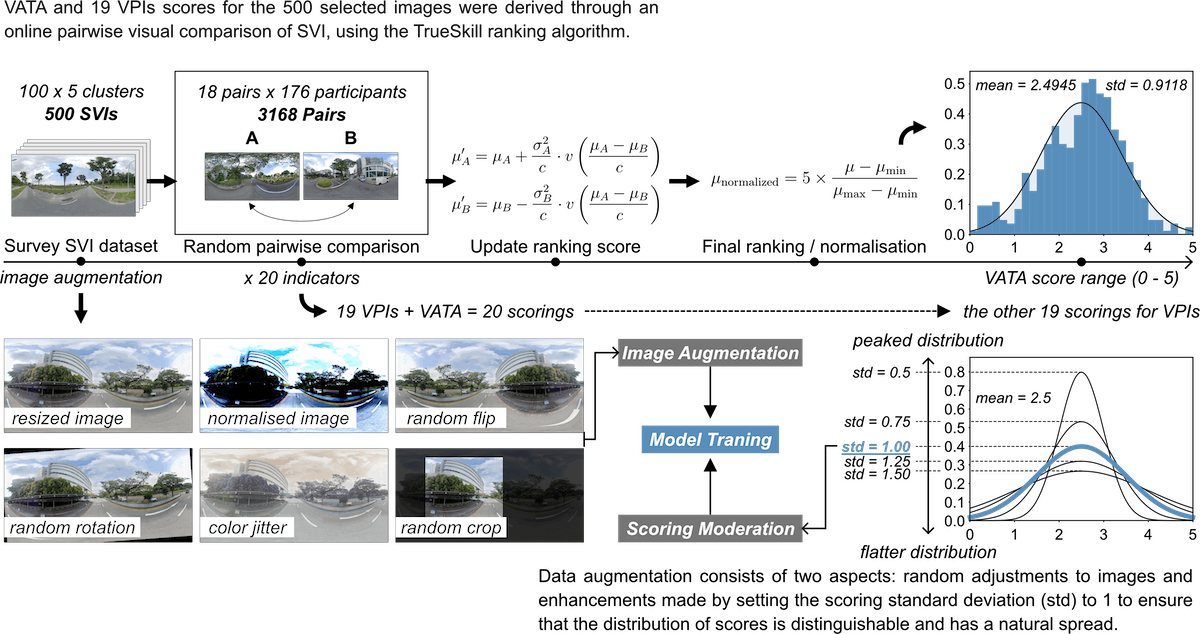} 
        \caption{Details on indicator scoring and data augmentation in the SVI visual assessment survey.} 
        \label{method_scoring}
    \end{figure}

    Along with VATA, the 19 VPIs are grouped into four categories: microclimate inference, built environment evaluation, streetscape design quality, and evoked emotion (Table \ref{method_question}). These VPIs were chosen to capture participants’ holistic visual-perceptual impressions of streetscapes, rather than having them rate individual streetscape elements. Participants were presented with system-randomised pairs of SVIs from the online survey for each indicator, which they compared directly. This approach encouraged a more natural expression of their perceptions, free from the pressure of explicit grading tasks. The TrueSkill algorithm \citep{herbrich2006trueskill} was subsequently employed to analyse responses, generating rankings and scores for the 500 SVIs on VATA and each VPI. These scores were then normalised to a 0-5 scale for machine learning regression (Figure \ref{method_scoring}). The ranking algorithm consistently updates the ranking score of winners and losers from pairwise comparisons, utilising the survey results comprising 3168 pairs collected from all the 176 participants for scoring calculation. Since one assumption of the TrueSkill algorithm is that the performance of all ranking elements (500 images in this case) follows a normal distribution, we can ultimately visualise the scoring results as a normal distribution within a range of 0 to 5 after normalisation, with the ideal population mean of 2.5. The TrueSkill process was repeated 20 times to obtain scores for VATA and each VPI. Details of the TrueSkill algorithm and survey results are introduced in \ref{appendix_trueskill} and \ref{appendix_survey}. These TrueSkill-derived scores enable us to employ predictive and inferential modelling to construct the IF–VPI–VATA relationship.

    To enhance model training based on SVIs and corresponding indicator scores from the online SVI visual assessment survey, we apply six data augmentations—resizing, normalising, random clipping, rotating, colour jittering, and cropping—to the 500 images. These transformations enhance the model’s generalisation and robustness (Figure \ref{method_scoring}). For VATA and VPI scores, we set the standard deviation (std) of each scoring distribution to 1, balancing discriminative power and an even score distribution. As illustrated in Figure \ref{method_scoring}, a std of 0.5 centralises the scores, limiting range, while a std of 1.5 flattens the distribution, reducing distinction. Setting std to 1 strikes an optimal balance between clarity and evenness.

    \begin{figure}[!ht]
        \centering    
        \includegraphics[width=1\textwidth]{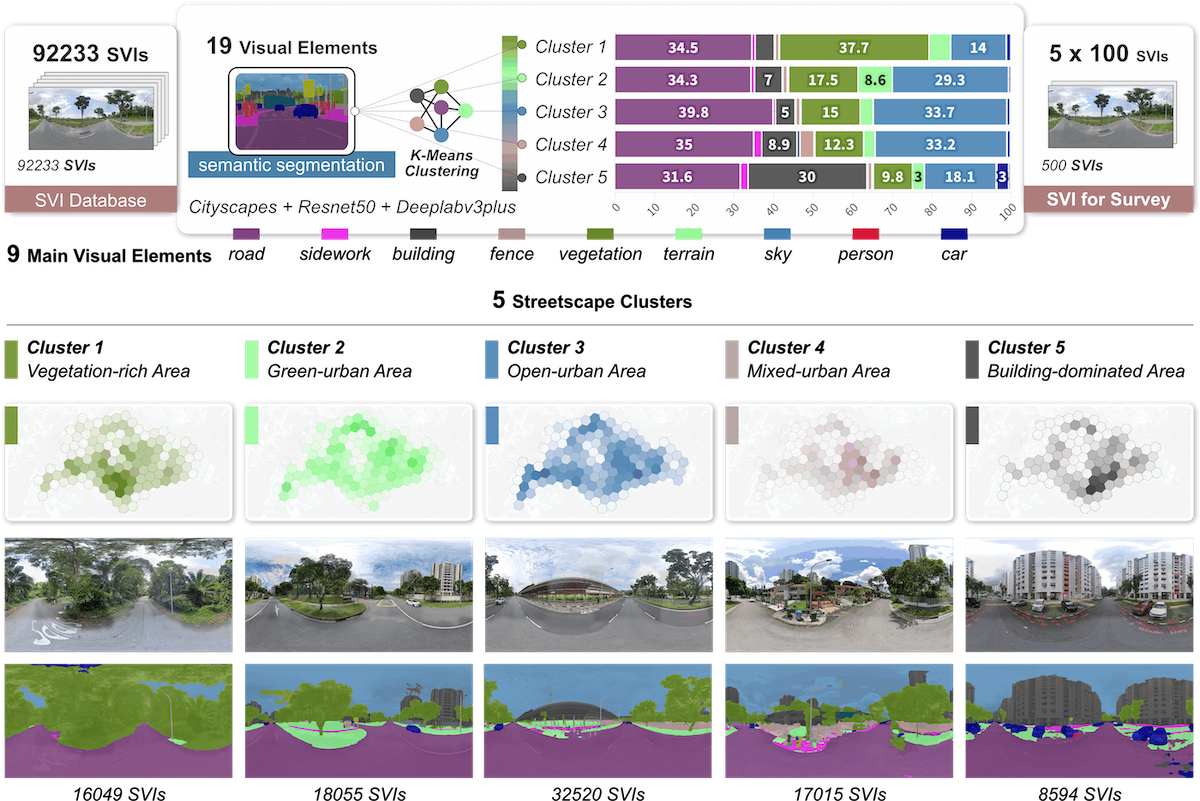} 
        \caption{SVI clustering based on segmentation results, and selection process for 500 SVIs taken for the online SVI visual assessment survey.} 
        \label{method_svi selection} 
    \end{figure}

    To ensure diverse streetscape representation, 500 SVIs were selected from a dataset of 92,233 SVIs across Singapore through the following process:
    
    \begin{enumerate}
        \item $k$-means clustering of all SVIs based on 19 streetscape elements from semantic segmentation, following the streetscape classification method outlined by \cite{liang2023revealing}.
        \item Categorisation into five types of streetscape: vegetation-rich, green-urban, open-urban, mixed-urban, and building-dominated.
        \item Random selection of 100 images from each category, totalling 500 images for the survey.
    \end{enumerate}

    Figure \ref{method_svi selection} illustrates the streetscape element proportions and spatial distributions of these categories. This balanced selection ensures equitable representation of each streetscape type, enhancing the VATA model's effectiveness by encompassing diverse urban environments. This approach underscores the importance of diversity and representativeness in training data for machine learning applications in complex urban settings.

\subsection{IF extracted from SVI data for model training and prediction}
    SVI data captures key aspects of urban street environments, such as greening, architectural style, and overall urban scenes, which features delineate the objective streetscape characteristics that contribute to thermal affordance. By combining IFs from 500 SVIs with visual assessment survey results, we train a MTNNL to model 19 VPIs and VATA effectively.
    
    As illustrated in Table \ref{method_datasets}, five categories of IF can be extracted from a single SVI $i$ using various algorithms and datasets: semantic segmentation identifies streetscape elements (e.g., buildings, vegetation), convolutional data capture high-dimensional pixel representations, object recognition counts objects (e.g., people, cars), pixel feature analysis examines colour features, and scene recognition assesses the likelihood of displaying specific scene type (e.g., commercial, natural). Pre-trained models like DeepLabV3+ \citep{chen2018encoder}, ResNet-50 \citep{he2016deep}, Faster R-CNN \citep{ren2015faster}, and OpenCV \citep{bradski2000opencv}, along with datasets like Cityscape \citep{cordts2016cityscapes}, COCO \citep{lin2014microsoft}, and Places365 \citep{zhou2017places}, are used to extract these IFs.
    
    Each SVI ($h \times w$ pixels) undergoes feature extraction, yielding five tensors concatenated into a single tensor $X = (P, F, A, C, S)$. For a dataset of $m$ images, this forms a tensor $D_1$ of size $N \times m$, where $N$ is the total number of features in $X_i$. We also create dataset tensor $D_2$ for an ENRM-based VATA inference model, excluding convolutional features for better interpretability. Thus, this study provides two IF datasets for modelling: $D_1 = [P_i, F_i, A_i, C_i, S_i]$ and $D_2 = [P_i, A_i, C_i, S_i]$ for \(i = 1, \dots, m\). $D_1$ serves as input for the MTNNL-based VATA prediction model, while $D_2$ trains the ENRM-based VATA inference model, showing interpretable correlations between IF, VPI, and VATA. Sub-features of IF under each category are listed in \ref{appendix_subfeatures}.

    \begin{table}[!ht]
        \caption{Models and datasets applied for IF extraction from SVI data.}
        \label{method_datasets}
        \centering
        \footnotesize
        \begin{tabularx}{\textwidth}{lllXl}
            \toprule
            \textbf{Extraction Method} & \textbf{Model/Lib} & \textbf{Dataset} & \textbf{Symbol} & \textbf{Feature Extracted} \\
            \midrule
            Semantic Segmentation & DeepLabV3+ & Cityscape & $P_i$ & 19 streetscape elements \\
            Convolutional Data & ResNet-50 & - & $F_i$ & convolutional tensor \\
            Object Detection & Fast R-CNN & COCO & $A_i$ & 91 object categories \\
            Pixel Feature & OpenCV & - & $C_i$ & 12 pixel visual feature  \\
            Scene Recognition & ResNet-50 & Places365 & $S_i$ & 365 scene categories \\
            \bottomrule
        \end{tabularx}
    \end{table}

\subsection{Architecture of VATA prediction and inference model}
    VATA employs two main model architectures: the multi-task neural network learning (MTNNL) for predictive analysis and the elastic net regression model (ENRM) for inference analysis. MTNNL is used for predicting VATA outcomes based on SVI and human visual assessment results, leveraging its strong predictive power. ENRM, as a linear and interpretable model, addresses collinearity issues and is suited for VATA inference research. With separate scores obtained for IFs, VPIs, and VATA for each image in the sample, we can use MTNNL to predict VPIs and VATA scores from IF data, and employ ENRM to quantitatively determine how the visual environmental features encapsulated in IFs influence both VPIs and VATA. In this process, VPIs act as an intermediate layer that not only improves VATA prediction accuracy but also deepens our understanding of the IF–VPI–VATA relationships.

    \begin{figure}[!ht]
        \centering    
        \includegraphics[width=1\textwidth]{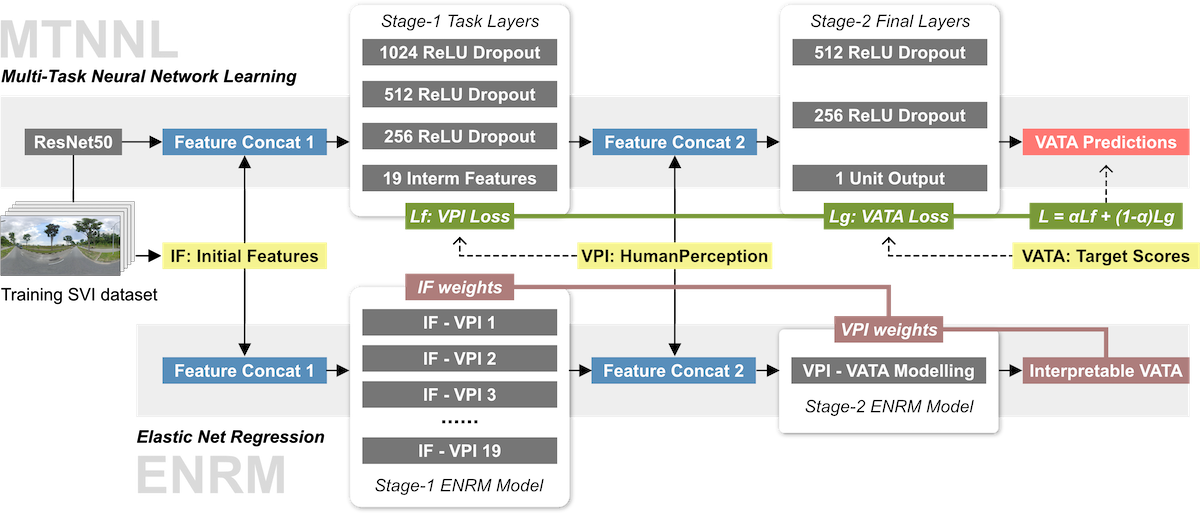} 
        \caption{Overall architecture of VATA prediction and inference modelling} 
        \label{method_model arch} 
    \end{figure}

\subsubsection{MTNNL Model for VATA prediction}
A MTNNL model is employed as the VATA prediction model and is trained in two simultaneous stages, as shown in Figure \ref{method_model arch}. The model architecture is designed to allow for concurrent learning of two related tasks, each corresponding to a different stage of training. In the first stage, the model is trained on the dataset tensor $D_1$, which contains five types of IFs in $m$ SVI images. Simultaneously, an intermediate tensor $V$ is created, representing human visual-perception rating data for the same $m$ images, where each image is graded on a 0-5 scale across 19 human VPIs derived from our online SVI visual assessment survey.
\begin{equation}
D_1 = 
    \begin{bmatrix}
    \mathbf{x}_1 &
    \mathbf{x}_2 &
    \cdots &
    \mathbf{x}_m 
    \end{bmatrix}, 
V = 
    \begin{bmatrix}
    \mathbf{v}_1 &
    \mathbf{v}_2 &
    \cdots &
    \mathbf{v}_m 
    \end{bmatrix}
\end{equation}
Here, $\mathbf{x}_i$ is an $n$-dimensional feature vector for the $i$-th image, and $\mathbf{v}_i$ is a 19-dimensional perception vector for the same image. The model learns a function $f: D_1 \rightarrow V$ mapping the image feature space to the human visual perception indicator space.

In the second stage, $D_1$ and $V$ are combined to predict VATA scores, also on a 0-5 scale, learning a function $g: (D_1, V) \rightarrow Y$, where $Y$ represents the VATA tensor for $m$ images. The neural networks for these two stages are trained simultaneously, and overall learning process involves minimizing the following integrated loss function:
\begin{equation}
L = \alpha L_f + (1 - \alpha) L_g
\end{equation}
where $L_f$ is the loss for predicting the VPI tensor $V$ from $D_1$, and $L_g$ is the loss for predicting VATA from $D_1$ and $V$. The parameter $\alpha$ controls the balance between the two tasks.

During the training phase, the visual IF data derived from 500 images across five feature extraction aspects, along with the VPI perceptual data across 19 metrics and the final VATA data, will be partitioned into a training set, a validation set, and a test set, adhering to a ratio of 60\%, 20\%, and 20\%, respectively. The model's final predictive performance will be assessed based on its MAE, MSE, RMSE, and adjusted $R^2$ on the test set, while the validation set plays a critical role during the training process by helping to tune hyperparameters, select the best model configuration, and monitor overfitting. The details of the model training data and process are presented in \ref{appendix_training}. 

\subsubsection{ENRM Model for VATA inference}
The ENRM model is used to establish an interpretable relationship between IF in SVI data, human VPI, and VATA in our studies. Combining ridge (L2) and lasso (L1) regularizations, ENRM effectively handles feature collinearity and performs variable selection. The model coefficients indicate the importance of each feature in explaining VATA scores.

Given a dataset tensor $D_2$ containing visual feature data from $m$ images, the ENRM can be mathematically formulated as:

\begin{equation}
\hat{y}_i = \mathbf{x}_i^\top \beta \quad,\quad \|\beta\|_1 = \sum_{j=1}^{p} |\beta_j| \quad,\quad \|\beta\|_2^2 = \sum_{j=1}^{p} \beta_j^2
\end{equation}

where $\hat{y}_i$ is the VATA score for the $i$-th image, $\mathbf{x}_i$ is the feature tensor of the $i$-th image, $\beta$ is the coefficient tensor. Combining these elements, the ENRM can be mathematically formulated as:

\begin{equation}
\min_{\beta} \left\{ \frac{1}{2m} \sum_{i=1}^{m} \left( y_i - \mathbf{x}_i^\top \beta \right)^2 + \lambda_1 \|\beta\|_1 + \lambda_2 \|\beta\|_2^2 \right\}
\end{equation}

where $\lambda_1$ and $\lambda_2$ are regularisation parameters for L1 and L2 penalties, respectively.

The L1 regularisation component helps in feature selection by shrinking irrelevant feature coefficients to zero, simplifying the model. The L2 regularisation component addresses multicollinearity among features, ensuring stable coefficient estimates. This model provides interpretable insights into the two-stage inference relationship between IF, VPI, and VATA scores (Figure \ref{method_model arch}). By identifying key visual features and perceptions contributing to VATA in urban settings, the ENRM aids in developing informed urban design strategies.

\subsection{VATA validation based on OTC dataset}

    To validate the effectiveness of the VATA framework in assessing human thermal comfort in urban street environments, we employed an on-the-ground OTC field survey dataset and data-driven statistical evaluations. The validation utilised OTC measurements from field investigations conducted on the campus of the National University of Singapore (NUS)  \citep{liu2023towards}. These field investigations were carried out independently from both the SVI-based visual assessments and the subsequent VATA modelling efforts, ensuring that there was no pre-established alignment or dependency between the datasets. This relative independence enhances the robustness of the validation process by minimizing the risk of bias and reinforcing the credibility of the comparative analysis.

\begin{figure}[!ht]
        \centering    
        \includegraphics[width=1\textwidth]{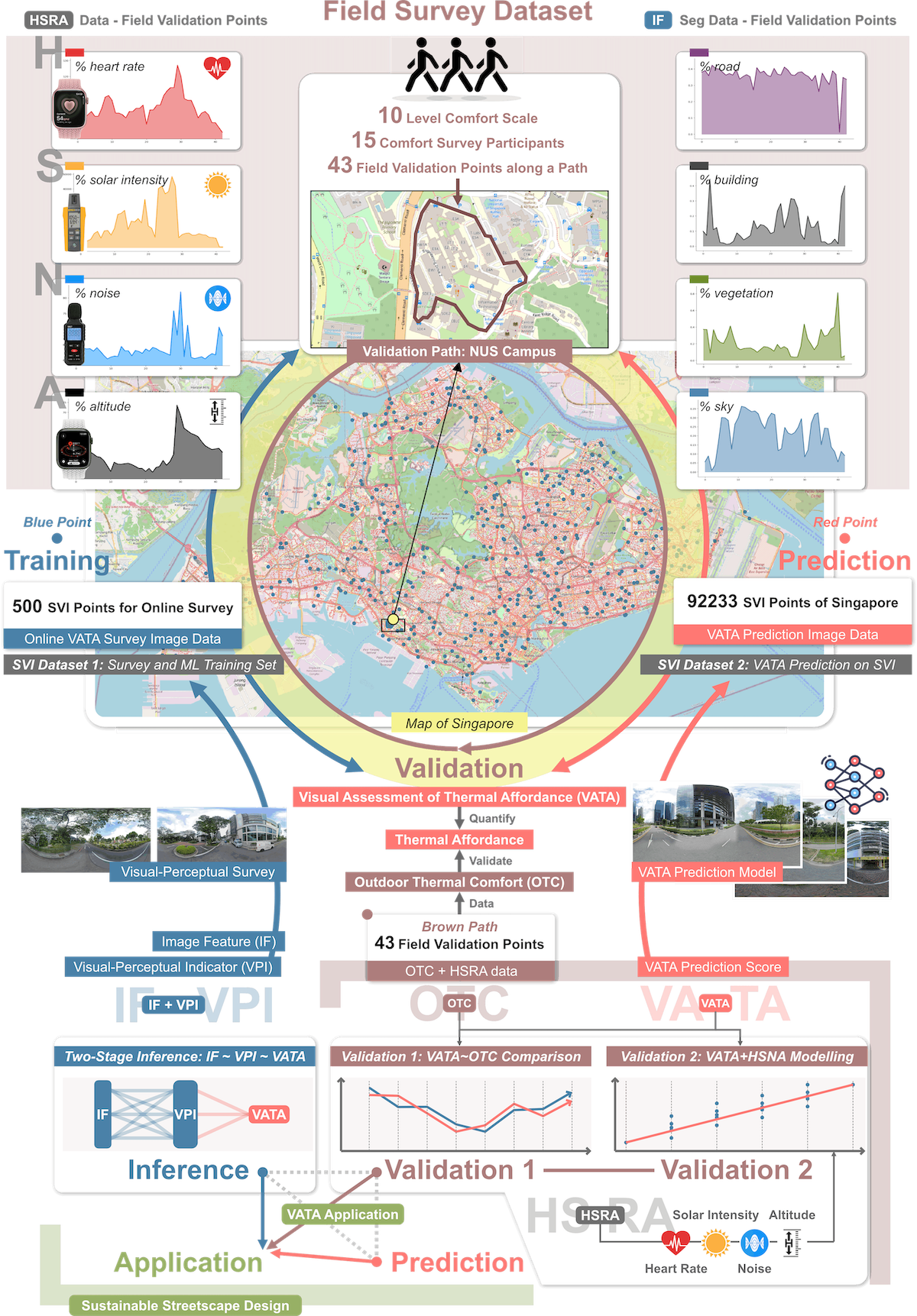} 
        \caption{Details of applying OTC field survey dataset as a validation method for VATA framework} 
        \label{method_validation} 
    \end{figure}

    The survey involved 15 NUS students and 43 investigation points along a continuous path (Figure \ref{method_validation}). The participants walked the path, providing comfort ratings (1-10 scale) at each point. They wore Apple Watches with the Cozie app \citep{tartarini2023cozie} to collect subjective feedback and heart rate data. Sound and solar intensity were measured using a sound metre and a smartphone, respectively. These 'thermal walk' experiments \citep{2024_3dgeoinfo_thermal_walk} were carried out three times daily (10 am, 2 pm, 5 pm) over five consecutive days in July 2022. The resulting OTC dataset includes average comfort scores, heart rate values, sunlight intensity, noise levels, and altitude measurements. We specifically focused on these parameters—readily captured via portable devices such as smartphones and wearable sensors—due to their potential for rapid variability across short distances. This enables fine-grained assessment of thermal comfort differences along the walking route, without the added complexity of factors like humidity or air temperature that tend to remain relatively uniform over these small spatial scales. While including additional factors such as humidity or air temperature measurements could enhance the comprehensiveness of comfort modelling, doing so is not the central focus of this study. Additionally, although Singapore provides publicly available air temperature data, the uneven spatial distribution and limited number of weather stations (where sensors are located) create significant uncertainties when interpolating localized conditions. Furthermore, the lack of measurements at very short intervals further reduces the granularity of these data for our fine-scale analysis.

    In addition, Google Street View images corresponding to each of the 43 investigation points were collected to facilitate visual-based VATA predictions and validations. The Google Street View imagery used was captured in the same year as the field investigation (2022), thus ensuring temporal synchronization. This alignment in timing allowed for a more reliable comparison and modelling between the VATA prediction scores and the on-site OTC measurements. The validation methodology, described in Figure~\ref{method_validation}, comprises two parts: 1) comparing VATA with OTC to assess thermal affordance effectiveness by directly comparing SVI-predicted VATA at 43 data points with corresponding OTC values; and 2) using a multivariate linear modelling approach combining VATA and HSNA indicators (heart rate, solar intensity, noise level, and altitude) to evaluate VATA reliability in predicting OTC.

    \clearpage

\section{Results}
\subsection{VATA prediction model performance}

    \begin{table}[!ht]
        \caption{VATA prediction model performance.}
        \label{result_model performance}
        \centering
        \footnotesize
        \begin{tabularx}{\textwidth}{lXXXX}
            \toprule
            \textbf{Model} & \textbf{MAE} & \textbf{MSE} & \textbf{RMSE} & \textbf{adjusted $R^2$}\\
            \midrule
            Decision Tree & 0.4509 & 0.7430 & 0.8620 & 0.1357 \\
            KNN Regression & 0.4192 & 0.6340 & 0.7963 & 0.2625 \\
            Support Vector Regression & 0.3929 & 0.5858 & 0.7654 & 0.3186 \\
            \textbf{ENRM (Lasso + Ridge)} & \textbf{0.2402} & \textbf{0.3159} & \textbf{0.5621} & \textbf{0.6325} \\
            LightGBM & 0.2422 & 0.3106 & 0.5573 & 0.6388 \\
            Random Forest & 0.2161 & 0.2919 & 0.5403 & 0.6604 \\
            Extreme Gradient Boosting & 0.2282 & 0.2906 & 0.5391 & 0.6619 \\
            CatBoost & 0.2260 & 0.2888 & 0.5374 & 0.6640 \\
            \textbf{MTNNL} & \textbf{0.1960} & \textbf{0.2339} & \textbf{0.4836} & \textbf{0.7316} \\
            \bottomrule
        \end{tabularx}
    \end{table}

    The MTNNL model has demonstrated superior performance in predicting VATA on the test set. As presented in Table \ref{result_model performance}, a comparative analysis was conducted against various models, including Decision Tree, KNN Regression, Support Vector Regression, ENRM, LightGBM, Random Forest, Extreme Gradient Boosting, and CatBoost. The results show that MTNNL consistently outperforms these models, achieving the highest adjusted $R^2$ (0.7316) and the lowest MAE (0.1960). This improved performance is attributed to the model’s two-stage, two-task prediction architecture, which leverages the advantages of deep learning for complex prediction tasks. Instead of directly predicting VATA scores from IF, the MTNNL model incorporates an intermediary variable layer, VPI, to account for human visual assessments, as shown in Figure \ref{method_framework}. This two-stage approach enhances the model’s ability to accurately predict VATA scores.

\subsection{VATA framework validation results}
\subsubsection{SVI-predicted VATA score vs. surveyed comfort data}

    \begin{figure}[!ht]
        \centering    
        \includegraphics[width=1\textwidth]{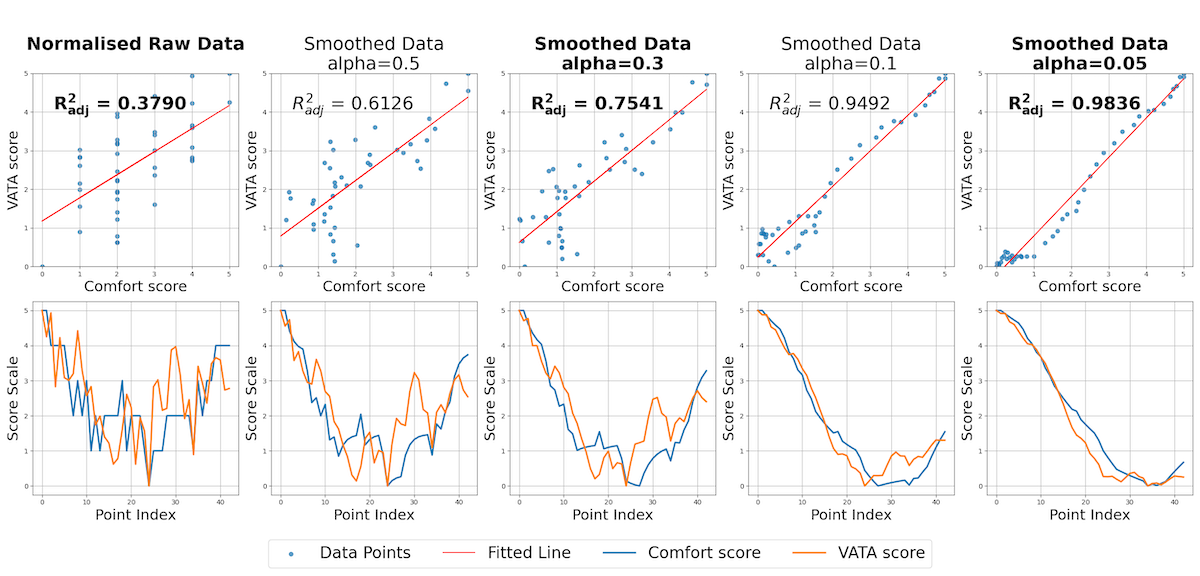} 
        \caption{VATA predictions fitting surveyed thermal comfort data along the path ($p<0.001$).} 
        \label{result_validation_fitting_figure} 
    \end{figure}
    
    Using our OTC field survey dataset, we analysed the relationship between SVI-predicted VATA scores and surveyed subjective comfort data across 43 discrete points along a path. To account for non-contiguous data points, we applied an exponential moving average (EMA) to smooth the VATA and OTC data, with the smoothing formula defined as:
        \begin{equation}
        S_i = \alpha \cdot X_i + (1 - \alpha) \cdot S_{i-1}
        \end{equation}
    where \( S_i \) is the smoothed value at the \( i \)-th data point, \( \alpha \) is the smoothing factor (between 0 and 1), \( X_i \) is the actual value at the \( i \)-th data point, and \( S_{i-1} \) is the smoothed value at the \( i-1 \)-th data point. Figure \ref{result_validation_fitting_figure} present the results for different smoothing factors.

     The raw data yielded an adjusted $R^2$ of 0.3790, indicating that VATA explained 37.90\% of the variance in comfort data. As the smoothing factor decreased from $0.5$ to $0.05$, the adjusted $R^2$ steadily improved, approaching 1, indicating a stronger fit; this suggests that our VATA prediction effectively captures the trend in the comfort data. Compared to the smoothing factor of 0.05, $\alpha$ equal to 0.3 can be a more balanced smoothing factor in VATA evaluation practice, which considers both continuous trends and actual measurement values, with an adjusted $R^2$ of 0.7541. These results confirm the validity of VATA as a significant predictor of in-field OTC variations, though additional factors like physiological parameters and sunlight intensity also influence thermal comfort.

\subsubsection{VATA-assisted comfort modelling vs. IF-assisted comfort modelling}
\nomenclature[B]{HSNA}{heart rate, solar intensity, noise level, altitude}

    \begin{figure}[!ht]
        \centering    
        \includegraphics[width=1\textwidth]{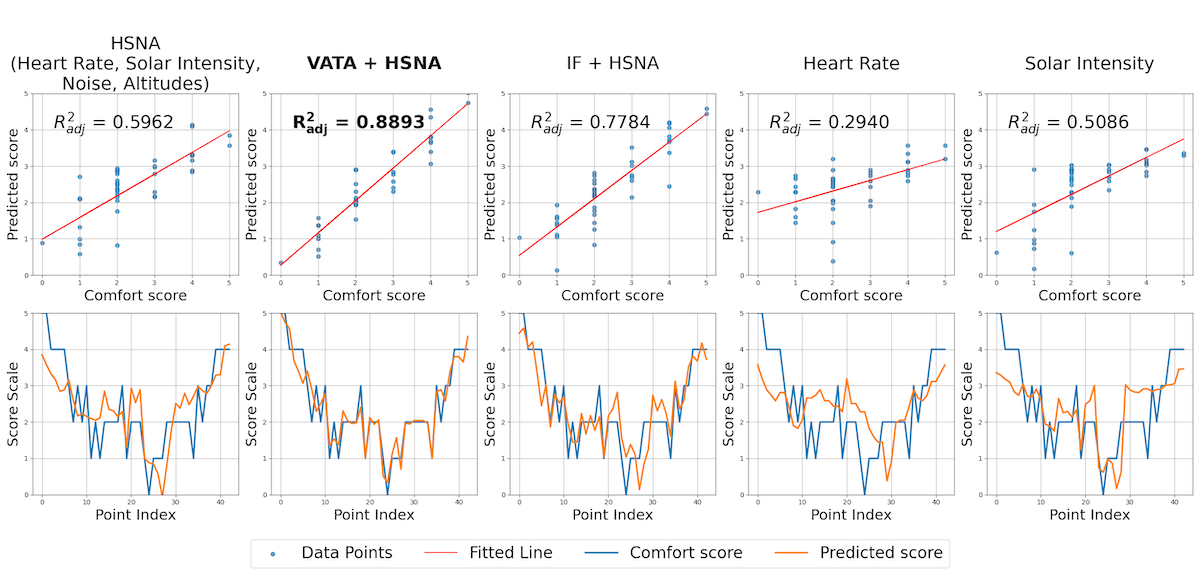} 
        \caption{Multivariate comfort modelling performance based on HSNA, VATA, and IF data ($p<0.001$).} 
        \label{result_validation_modelling_figure} 
    \end{figure}

    We analysed the 43-point OTC field survey comfort data using HSNA parameters (heart rate, solar intensity, noise level, altitude). The performance of VATA+HSNA, IF+HSNA, and individual heart rate and solar intensity models is summarised in Figure \ref{result_validation_modelling_figure}. The VATA+HSNA model outperformed all others, achieving an adjusted $R^2$ of 0.8893, surpassing IF+HSNA (adjusted $R^2 = 0.7784$). This highlights VATA’s superior ability to model thermal comfort compared to IF-based models. One of the potential reasons for VATA’s improved performance can be its ability to incorporate human visual-perceptual evaluations to construct the perception logic chain IF-VPI-VATA-OTC, which more accurately capture comfort perceptions.
    
    HSNA alone explained 59.62\% of the variance in comfort data, with heart rate and solar intensity models achieving adjusted $R^2$ values of 0.2940 and 0.5086, respectively. While VATA is a strong explanatory variable for thermal comfort, accurate prediction of OTC requires its integration with HSNA and additional environmental factors.

\subsection{VATA mapping and spatial analysis}

    \begin{figure}[!ht]
        \centering    
        \includegraphics[width=1\textwidth]{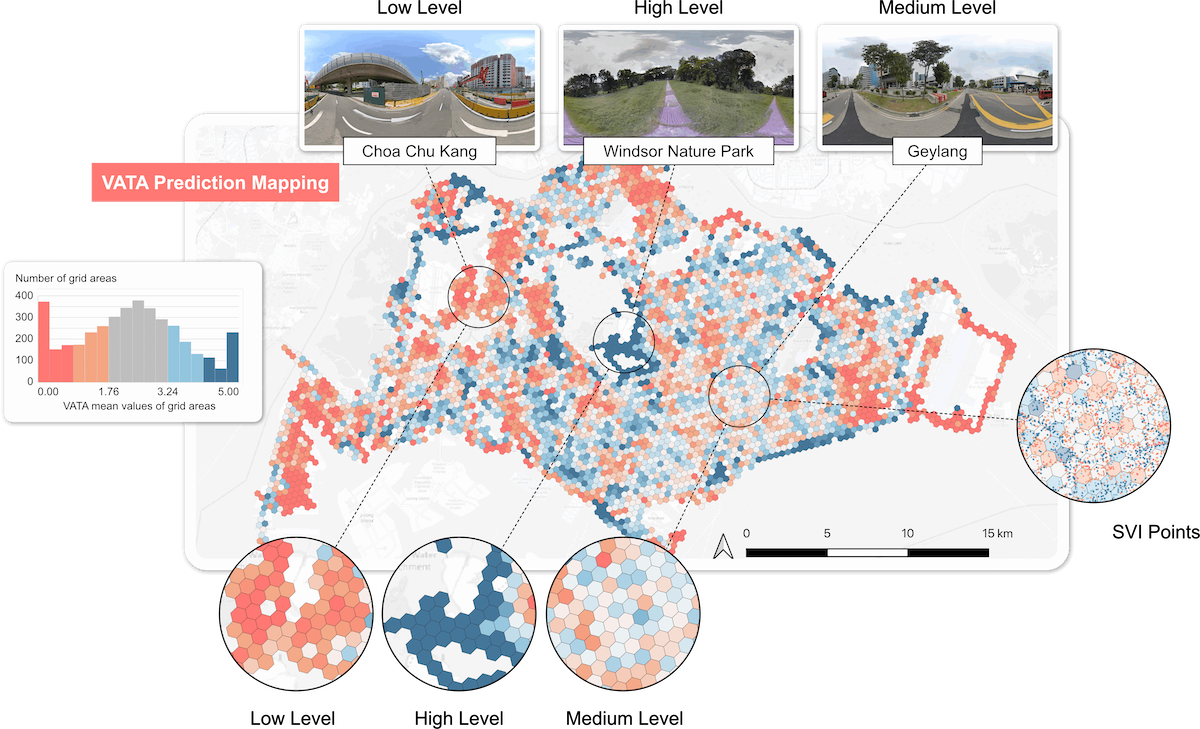} 
        \caption{VATA prediction result mapping for Singapore streetscapes.} 
        \label{result_VATA mapping} 
    \end{figure}

    \begin{figure}[!ht]
        \centering    
        \includegraphics[width=1\textwidth]{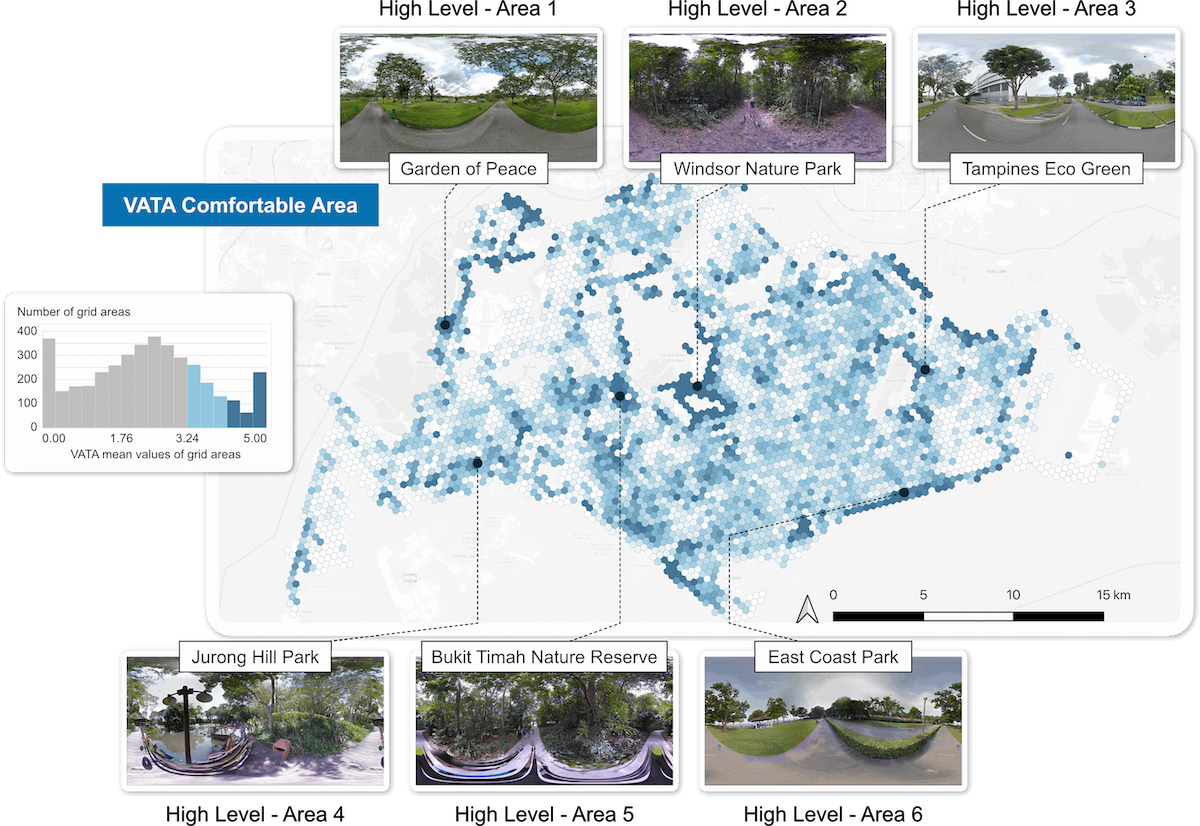} 
        \caption{High level VATA area: Spatial distribution of VATA-comfortable streetscape area} 
        \label{result_VATA mapping_good} 
    \end{figure}
    
    \begin{figure}[!ht]
        \centering    
        \includegraphics[width=1\textwidth]{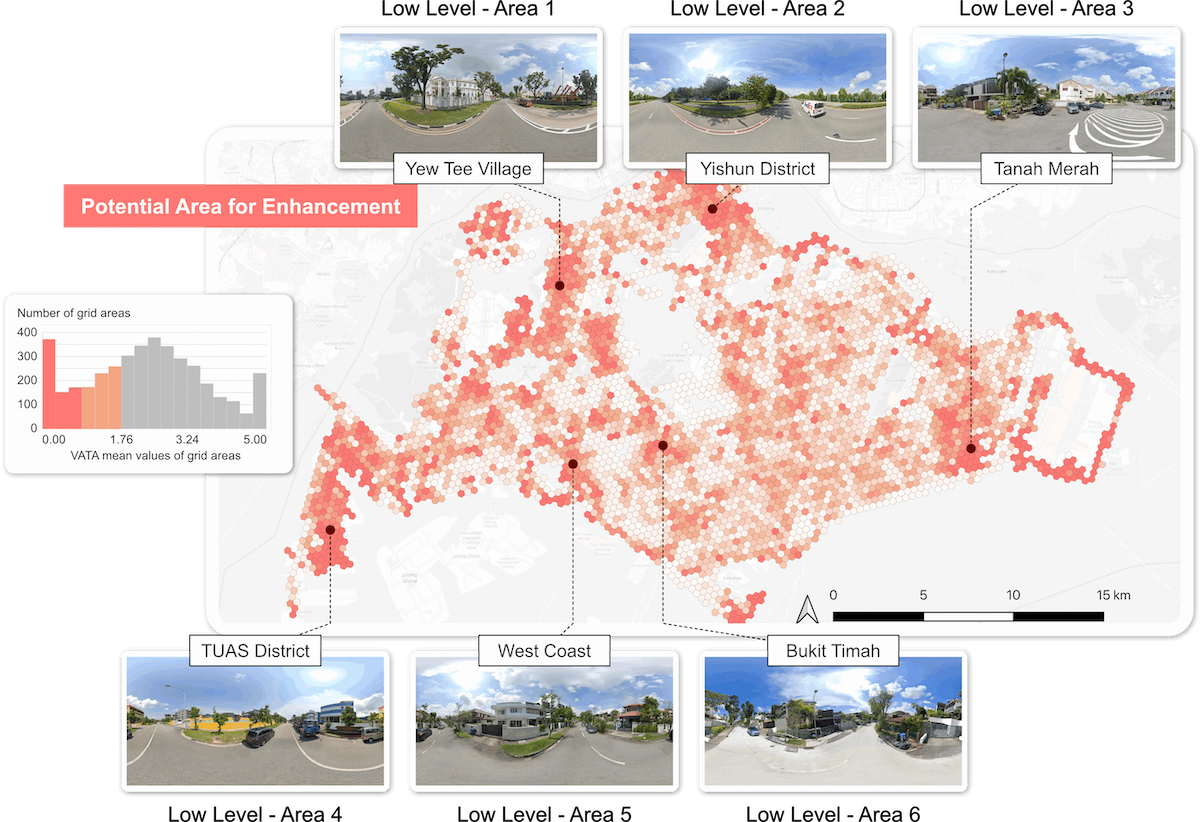} 
        \caption{Low level VATA area: Spatial distribution of potential area for streetscape enhancement} 
        \label{result_VATA mapping_bad} 
    \end{figure}

    The trained VATA prediction model assigns a VATA value to each of the 92,233 SVIs and aggregates them as average scores within hexagonal spatial units (Figure~\ref{result_VATA mapping}). Using the H3 geospatial indexing system at resolution level 9, each hexagonal unit covers approximately 0.1 square kilometres, balancing precision and scalability for urban planning-orientated geospatial analytics. The VATA framework offers a valuable tool for enhancing urban quality of life by informing sustainable streetscape planning and design. It provides a comprehensive analysis of thermal affordance at the urban scale. By calculating and visually representing the average VATA values for each hexagonal spatial cell, the model effectively communicates the visual evaluation of thermal affordance across urban streetscapes in Singapore. Hexagonal spatial cells are colour coded to reflect varying levels of thermal affordance: red indicates high VATA values ($3.24 < \text{VATA} \leq 5$), suggesting superior thermal conditions due to shading and vegetation; blue represents low values ($0 \leq \text{VATA} \leq 1.76$), indicating poor conditions from sun exposure and limited greenery; grey signifies moderate thermal affordance ($1.76 < \text{VATA} \leq 3.24$).
    
    The VATA mapping presented in Figure \ref{result_VATA mapping} effectively visualises thermal affordance across urban streetscapes. Low VATA areas, such as Choa Chu Kang, lack continuous walking paths, greenery, and shaded spaces, resulting in increased sun exposure and reduced comfort. In contrast, high VATA areas like Windsor Nature Park feature well-defined pedestrian paths and abundant vegetation, offering better thermal affordance. Medium-rated areas such as Geylang neighbourhood provide moderate comfort. Despite continuous pedestrian paths and shaded areas, proximity to high-traffic zones or dense housing areas introduces additional artificial heat sources that diminish overall thermal affordance. 

    Figures \ref{result_VATA mapping_good} and \ref{result_VATA mapping_bad} illustrate the spatial distribution of high and low VATA values, respectively. High-VATA areas, particularly city parks such as Garden of Peace, Windsor Nature Park, and Bukit Timah Nature Reserve, offer superior thermal comfort due to abundant green spaces, highlighting their importance for urban planning and public health. The exceptional VATA performance of the area around East Coast Park further underscores the cooling effects of water bodies and greenery in urban heat mitigation. In contrast, Figure \ref{result_VATA mapping_bad} reveals low-VATA regions in the northwestern and eastern parts of Singapore, presenting opportunities for improving streetscape thermal affordance. Thus, the VATA framework serves as a valuable tool for urban planners and designers to identify and improve thermal affordance levels across different parts of a city, aiding the development of sustainable and liveable environments.

\subsection{IF-VPI-VATA two-stage inference modelling results}

    \begin{table}[!ht]
        \caption{ENRM models for IF-VATA, VPI-VATA, and IF+VPI-VATA inferences.}
        \label{result_inference model}
        \centering
        \footnotesize
        \begin{tabularx}{\textwidth}{l>{\centering\arraybackslash}X>{\centering\arraybackslash}X>{\centering\arraybackslash}X}
            \toprule
            \textbf{VATA Model} & \textbf{SVI Image Feature (IF)} & \textbf{Visual-Perceptual Indicator (VPI)} & \textbf{Two-stage Model (IF + VPI)} \\
            \midrule
            \multicolumn{4}{l}{\textbf{Model Parameter}} \\
            \cmidrule(lr){1-1} \cmidrule(lr){2-4}
            $\alpha$ & 0.0289 & 0.0049 & 0.0070 \\
            L1 Ratio & 0.5 & 0.5 & 0.5 \\
            Intercept & 2.4945 & 2.4945 & 2.4945 \\
            \midrule
            \multicolumn{4}{l}{\textbf{Performance}} \\
            \cmidrule(lr){1-1} \cmidrule(lr){2-4}
            R² & 0.6587 & 0.6918 & 0.7576 \\
            \textbf{Adjusted R²} & \textbf{0.6353} & \textbf{0.6823} & \textbf{0.7443} \\
            MAE & 0.2804 & 0.2633 & 0.2184 \\
            MSE & 0.2832 & 0.2557 & 0.2011 \\
            RMSE & 0.5322 & 0.5057 & 0.4485 \\
            MAPE(\%) & 24.9150 & 23.4235 & 18.9456 \\
            AIC & -564.8017 & -649.8209 & -747.8569 \\
            BIC & -425.7196 & -582.3871 & -634.0625 \\
            \midrule
            \multicolumn{4}{l}{\textbf{Validation}} \\
            \cmidrule(lr){1-1} \cmidrule(lr){2-4}
            Cross-Validation Score & 0.5000 & 0.4114 & 0.3816 \\
            Mean VIF & 2.4843 & 2.4923 & 2.0383 \\
            \midrule
            \multicolumn{4}{l}{\textbf{Model Complexity}} \\
            \cmidrule(lr){1-1} \cmidrule(lr){2-4}
            Significant Coefficients & 32 & 15 & 26 \\
            Total Features & 52 & 19 & 71 \\
            \bottomrule
        \end{tabularx}
    \end{table}
    
    After establishing the VATA prediction model using MTNNL, we developed an ENRM inference model based on 500 SVI-based online survey results to better understand the IF-VPI-VATA inference relationship, which helps future sustainable streetscape planning and design. Similarly to MTNNL, our ENRM-based inference modelling also employs a two-stage process: first, 52 IFs model 19 VPIs; second, the superposition of these two classes of features model VATA. The advantage of ENRM is that it efficiently addresses collinearity and identifies significant factors.

    Table \ref{result_inference model} presents the results for IF-VATA, VPI-VATA, and two-stage IF-VPI-VATA model. The two-stage model achieved the best performance with an adjusted $R^2$ of 0.7443, comparable to the MTNNL-based prediction model (adjusted $R^2$ = 0.7316). The feature weights derived from the two-stage IF–VPI–VATA model highlight the positive and negative factors influencing thermal affordance. A detailed discussion of these factors, including positive features (e.g., vegetation, shading area, human scale) and negative features (e.g., traffic elements, complexity, sunlight intensity), as well as design insights drawn from the inference modelling, is provided in \ref{appendix_inference}.

\section{Discussion}
    \subsection{VATA framework on sustainable streetscape planning and design}
    Our proposed VATA framework offers a novel approach for evaluating the thermal affordance of urban streetscapes at both urban-scale coverage and street-level resolution. Utilising SVIs across Singapore, computer vision algorithms to extract IF, and an online SVI visual assessment survey to investigate VPI, VATA supports sustainable streetscape design. The high-resolution urban-scale VATA maps, created after VATA prediction modelling and derived from averaging SVI scores within hexagonal spatial units, are instrumental in identifying urban areas with lower thermal affordance, helping urban planners prioritise regions for streetscape improvements and enhance sustainability. This targeted approach facilitates effective thermal affordance evaluations and informed urban planning decisions.
    
    Furthermore, the inference modelling in our VATA framework is particularly valuable for helping urban planners and designers understand objective streetscape compositions (through 52 IFs), individuals' subjective perceptions (through 19 VPIs), and their combined impact on thermal affordance. By analysing the inference relationship IF-VPI-VATA, urban planners and designers can strategically and organically design and arrange streetscape elements that not only meet thermal affordance standards but also elevate the overall quality of sustainable urban streetscape planning and design.
    
\begin{figure}[!ht]
        \centering    
        \includegraphics[width=1\textwidth]{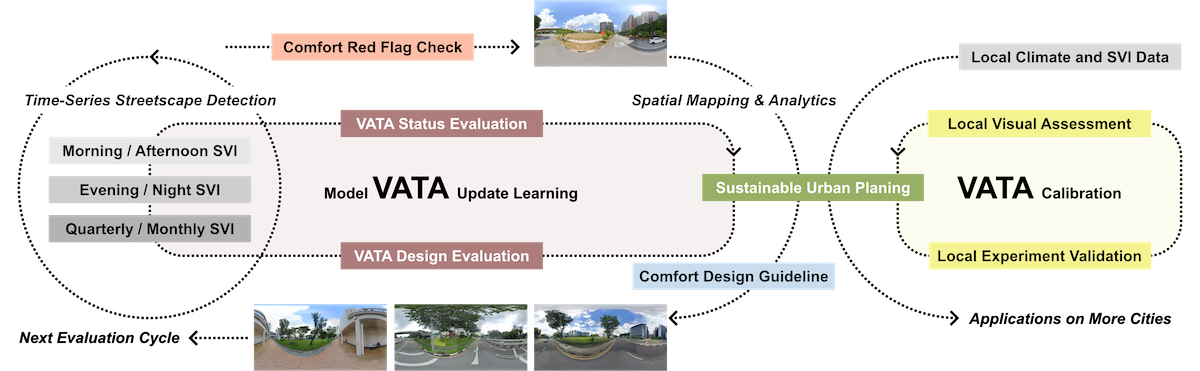} 
        \caption{Long-term sustainable streetscape planning and design based on continuous thermal affordance evaluation supported by VATA framework.} 
        \label{discussion_VATA framework} 
    \end{figure}

    Figure \ref{discussion_VATA framework} illustrates a continuous thermal affordance monitoring process and the corresponding sustainable streetscape planning and design supported by the evolving VATA framework. In each monitoring cycle, quarterly or monthly SVI data is collected, analysed using the VATA model, and further applied to develop the design guidelines to inform sustainable streetscape planning and design. This process updates the SVI dataset and VATA model for the next cycle, promoting a long-term sustainable and environmentally friendly streetscape planning and design paradigm in terms of thermal affordance.

    The potential applications of VATA extend further. For residents, VATA maps are instrumental in improving street-level thermal affordance navigation. For urban planners and researchers, the open source nature of the VATA framework encourages continuous improvement, integrating more diverse metrics for comprehensive urban thermal affordance analysis.

    Importantly, while our current research focuses on the hot tropical climate of Singapore, the VATA framework and its underlying workflow can be methodologically transferred to other cities. To achieve reliable outcomes under different climatic and urban conditions, it is essential to conduct new local SVI visual assessment surveys and incorporate locally relevant geospatial data. Moreover, performing on-site thermal comfort validations in the target cities would be a crucial step to fine-tune the model’s parameters, ensuring the VATA approach remains robust and accurate. By doing so, the VATA workflow can guide sustainable streetscape planning and design across diverse environments while maintaining methodological consistency, adaptability, and scientific rigour.

    \subsection{Limitations and future work}
    The primary objective of this research is to leverage visual features in SVI and human assessment for a cost-effective, scalable approach to thermal affordance evaluation. However, certain limitations suggest opportunities for future study. First, temporal variance in SVI was not considered, omitting the possible impacts of time, weather conditions, and seasons on thermal affordance assessment. Future research could address this by sampling street views from the same locations at various times. Second, the validation was conducted along a single continuous path with 43 investigation points, which limits the diversity of environmental conditions captured. Expanding to a broader selection of field locations and incorporating microclimatic parameters (e.g. humidity, air temperature, and wind speed) for the validation would enhance both model robustness and generalizability. 
    
    Additionally, while visual cues provide valuable insights, they represent only one facet of urban conditions. Future research might incorporate complementary data sources such as satellite imagery, remote sensing methods, or GIS-based analyses (e.g. land surface temperature or vegetation indices) to achieve a more comprehensive understanding of thermal affordance. Online surveys also pose limitations due to potential biases from participants’ varied backgrounds and limited professional expertise; however, they remain a practical means of data collection within manageable constraints.
    
    Finally, testing and validating this approach in multiple urban contexts, possibly spanning diverse climatic zones and geospatial conditions, would refine the framework and bolster its applicability. Through these efforts, future work can offer a more robust, scalable, and universally relevant tool for assessing and improving thermal comfort in urban environments worldwide.

\section{Conclusions and Contributions}
This research introduces the concept of thermal affordance and develops the visual assessment of thermal affordance -- VATA method to evaluate built environments' capability to enhance thermal comfort. Key conclusions and contributions include:
\begin{enumerate}
    \item Introducing thermal affordance, which describes the built environment’s hidden information and inherent capability to promote thermal comfort. This new concept represents objective and fixed environmental influences on thermal comfort perception.
    
    \item Pioneering the VATA framework, which integrates objective image features (IF) derived from street view imagery (SVI) with subjective visual-perceptual indicators (VPI) from survey data to build a two-stage neural network model for SVI-based cost-effective urban-scale thermal affordance assessment. 

    \item Exploring the IF–VPI–VATA relationship through inference modelling, contributing to the understanding of how visual compositions and human perceptions jointly shape thermal affordance, informing future sustainable streetscape planning and design to improve thermal comfort perception.

    \item Generating a spatially detailed urban-scale mapping of thermal affordance across Singapore's streetscape, which is validated by in-field thermal comfort data and transferable to other cities, prioritising regions for future sustainable streetscape renovation.

    \item Demonstrating that the validity of the VATA framework is supported by in-field comfort data. Comfort modelling based on VATA metrics outperforms models relying solely on IF metrics, highlighting the key role of VPI in the two-stage modelling process.

\end{enumerate}

\section*{Author Contributions}
\textbf{Sijie Yang:} Conceptualisation, Methodology, Investigation, Formal analysis, Software, Validation, Visualisation, Writing - Original Draft. \textbf{Adrian Chong:} Conceptualisation, Methodology, Writing - Review \& Editing. \textbf{Pengyuan Liu:} Data Curation, Resources, Methodology, Writing - Review \& Editing. \textbf{Filip Biljecki:} Conceptualisation, Methodology, Resources, Writing - Review \& Editing, Supervision, Project administration, Funding acquisition.

\section*{Data and Code Availability}
The computed VATA data for thermal affordance in Singapore can be accessed at \url{https://thermal-affordance.ual.sg/}. Data for additional cities will be updated regularly. The code and calculation methods will be open-sourced at \url{https://github.com/Sijie-Yang/Thermal-Affordance}.

\section*{Acknowledgements}
We gratefully acknowledge the participants of the experiment.
We thank our colleagues at the NUS Urban Analytics Lab for the discussions.
The research involves human participants. This experiment has been approved by the Ethical Review Committee (ERC) of the NUS College of Design and Engineering on 6 February 2024 (reference code 2024/DOA/002).
This research is supported by NUS Research Scholarship (NUSGS-CDE DO IS AY22\&L GRSUR0600042).
This research is part of the project Large-scale 3D Geospatial Data for Urban Analytics, which is supported by the National University of Singapore under the Start Up Grant R-295-000-171-133.
This research is part of the project Multi-scale Digital Twins for the Urban Environment: From Heartbeats to Cities, which is supported by the Singapore Ministry of Education Academic Research Fund Tier 1. 

\section*{Declaration of generative AI and AI-assisted technologies in the writing process}
We utilised Grammarly and Writefull for language refinement and grammar checks during the writing process. While the tools helped ensure language accuracy and clarity, all scientific insights, conclusions, and content were generated solely by the authors. The final manuscript was thoroughly reviewed and edited by the authors to maintain accuracy and integrity.

\printnomenclature

\appendix
\newpage
\section{Introduction on TrueSkill algorithm in this study}
\label{appendix_trueskill}

    The TrueSkill algorithm, developed by Microsoft Research, is used in this study to rank street view images (SVIs) based on pairwise comparisons made by survey participants. Each image's performance in a comparison is modelled as a normal distribution, where the observed performance \( p_i \) is the sum of its true skill level \( s_i \) and some random noise \( \epsilon_i \). This relationship can be expressed as follows:
    
    \[
    p_i = s_i + \epsilon_i \quad \text{where} \quad \epsilon_i \sim \mathcal{N}(0, \beta^2)
    \]
    Here, \( \epsilon_i \) is normally distributed with a mean of zero and variance \( \beta^2 \), representing uncertainties in human perception during the comparison process.
    
    After each comparison, the skill estimates of the two images involved are updated using Bayesian inference. If image \( i \) is rated higher than image \( j \), their skill levels are adjusted based on the following formulas:
    
    \[
    \mu'_i = \mu_i + \frac{\sigma_i^2}{c} \cdot \Delta, \quad \sigma_i'^2 = \sigma_i^2 \left( 1 - \frac{\sigma_i^2}{c^2} \right)
    \]
    Here, \( \mu'_i \) and \( \sigma_i'^2 \) are the updated mean and variance for the image \( i \), and \( \Delta \) represents the performance difference between the two images. The term \( c \) is a normalisation factor that accounts for the uncertainty in both images' skill estimates.
    
    TrueSkill operates under the assumption that all images' performances follow a normal distribution, which allows for an iterative refinement of the skill estimates through multiple pairwise comparisons. As more comparisons are made, the algorithm becomes more confident in the ranking, reflected in reduced variance for each image.
    
    Once all comparisons are completed, the raw TrueSkill scores are normalised to a 0-5 scale to ensure consistency across VATA and the 19 VPIs. The normalisation process is performed using the formula:
    
    \[
    S_i = \frac{s_i - \min(s)}{\max(s) - \min(s)} \times 5
    \]
    where \( S_i \) is the normalised score for image \( i \), and \( s_i \) is the raw skill score. This normalisation ensures that the scores are comparable across different indicators and suitable for further analysis, such as machine learning regression tasks.

\newpage
\section{Result of the online SVI visual assessment survey}
\label{appendix_survey}

\begin{figure}[!ht]
        \centering    
        \includegraphics[width=1\textwidth]{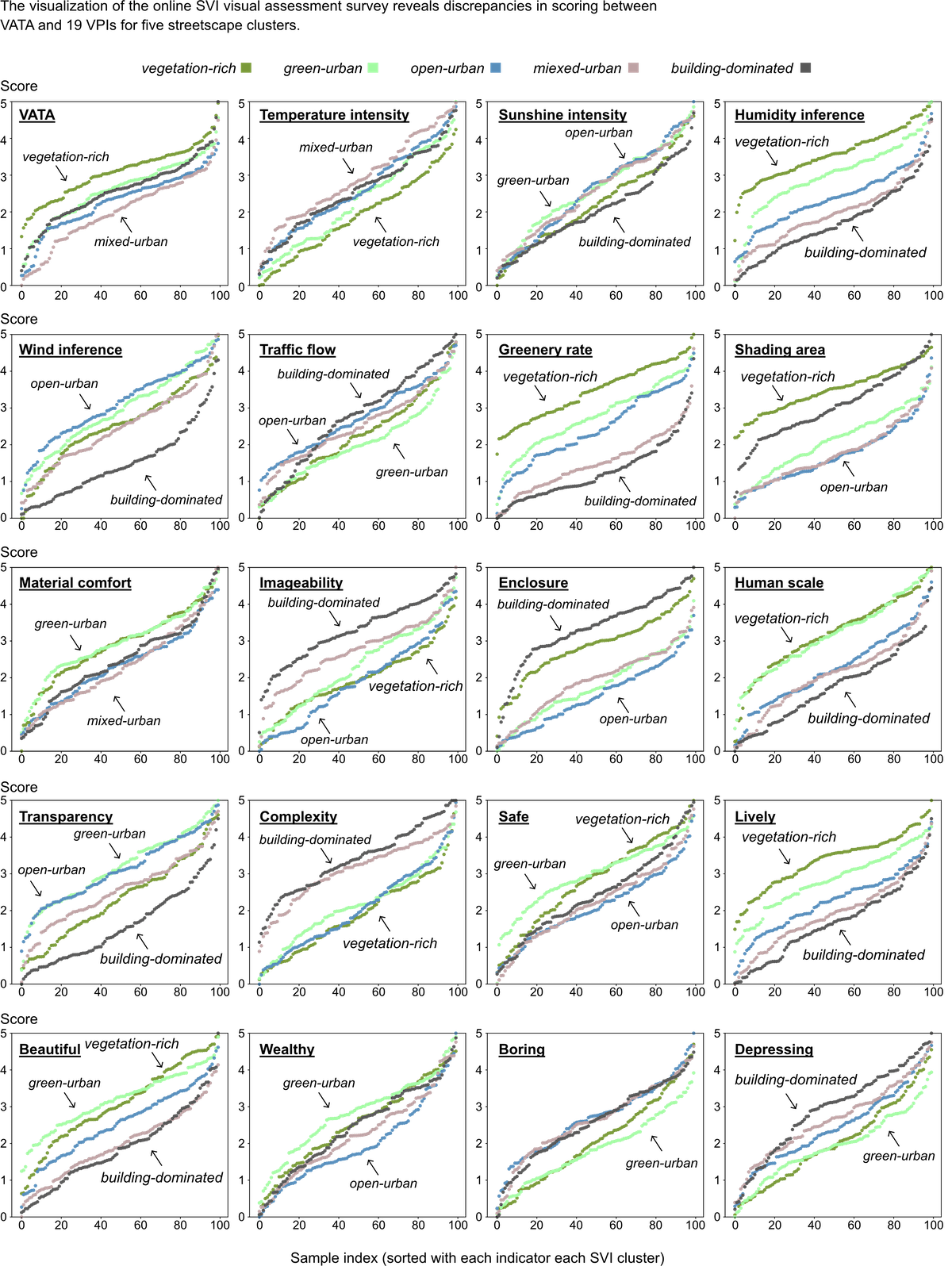} 
        \caption{Visualisation of online SVI visual assessment survey result.} 
        \label{appendix_survey result} 
    \end{figure}

    The results of the SVI visual assessment survey (Figure \ref{appendix_survey result}) reveal clear discrepancies in the scoring across all indicators, including VATA and the other 19 VPIs, when comparing the five streetscape clusters. Notable examples include indicators such as humidity inference, greenery rate, and imageability. Streetscapes rich in vegetation tend to receive higher average scores for humidity inference and greenery rate, whereas streetscapes dominated by buildings score higher on average for imageability. On the other hand, some indicators, like sunshine intensity and traffic flow, show much smaller discrepancies across the different clusters.

\newpage
\section{Subfeatures of IF datasets in VATA modelling}
\label{appendix_subfeatures}

    \begin{table}[!ht]
        \caption{Main sub-features applied for modelling and analytics.}
        \label{method_features}
        \centering
        \fontsize{6.5}{6.5}\selectfont%
        \begin{tabularx}{\textwidth}{p{2.6cm} p{0.3cm} p{1.7cm} >{\arraybackslash}p{6cm}}
            \toprule
            \textbf{Visual Feature Type} & {} & \textbf{Sub-feature} & \textbf{Definition} \\
            \midrule
            \textbf{Semantic Segmentation $P_i$} & & \\
                \cmidrule(lr){3-3} \cmidrule(lr){4-4} 
                    \phantom{aa} \multirow[t]{3}{*}{Built Env Element} & \multirow[t]{3}{*}{$P_{i-B}$} & $_{B1}$ building & The proportion of building structures like houses and offices.\\
                    {} & {} & $_{B2}$ wall & The proportion of wall, vertical barriers made of concrete.\\
                    {} & {} & $_{B3}$ fence & The proportion of fences, barriers made of wood or metal.\\
                \cmidrule(lr){3-3} \cmidrule(lr){4-4}
                    \phantom{aa} \multirow[t]{3}{*}{Transport Element} & \multirow[t]{3}{*}{$P_{i-T}$} & $_{T1}$ road & The proportion of roads, paved pathways for vehicular traffic. \\
                    {} & {} & $_{T2}$ traffic\_infra & The proportion of poles, traffic lights, and traffic signs.\\
                    {} & {} & $_{T3}$ vehicle &  The proportion of cars, trucks, buses, motorcycles, and trains. \\
                \cmidrule(lr){3-3} \cmidrule(lr){4-4}
                    \phantom{aa} \multirow[t]{3}{*}{Nature Element} & \multirow[t]{3}{*}{$P_{i-N}$} & $_{N1}$ sky & The proportion of sky in the image. \\
                    {} & {} & $_{N2}$ vegetation & The proportion of plants, trees, and bushes. \\
                    {} & {} & $_{N3}$ terrain & The proportion of terrain, ground features like soil or rocks. \\
                \cmidrule(lr){3-3} \cmidrule(lr){4-4}
                    \phantom{aa} \multirow[t]{3}{*}{Human-scale Element} & \multirow[t]{3}{*}{$P_{i-H}$} & $_{H1}$ person & The proportion of people walking or standing. \\
                    {} & {} & $_{H2}$ sidewalk & The proportion of pathways for pedestrians. \\
                    {} & {} & $_{H3}$ bicycle\_rider & The proportion of bicycles and riders. \\
                \cmidrule(lr){1-4}
            \textbf{Object Detection $A_i$} & & \\
                \cmidrule(lr){3-3} \cmidrule(lr){4-4} 
                    \phantom{aa} \multirow[t]{3}{*}{Infrastructure Object} & \multirow[t]{3}{*}{$A_{i-I}$} & $_{I1}$ traffic\_light & Number of traffic lights detected in the image.\\
                    {} & {} & $_{I2}$ stop\_sign & Number of stop signs detected in the image.\\
                    {} & {} & $_{I3}$ fire\_hydrant & Number of fire hydrants detected in the image.\\
                \cmidrule(lr){3-3} \cmidrule(lr){4-4} 
                    \phantom{aa} \multirow[t]{4}{*}{Vehicular Object} & \multirow[t]{4}{*}{$A_{i-V}$} & $_{V1}$ car & Number of cars detected in the image. \\
                    {} & {} & $_{V2}$ bus & Number of buses detected in the image.\\
                    {} & {} & $_{V3}$ truck & Number of trucks detected in the image. \\
                    {} & {} & $_{V4}$ motorcycle & Number of motorcycles detected in the image. \\
                \cmidrule(lr){3-3} \cmidrule(lr){4-4} 
                    \phantom{aa} \multirow[t]{3}{*}{Human-scale Object} & \multirow[t]{3}{*}{$A_{i-H}$} & $_{H1}$ person & Number of persons detected in the image. \\
                    {} & {} & $_{H2}$ bench & Number of benches detected in the image. \\
                    {} & {} & $_{H3}$ bicycle\_rider & Number of bicycles and riders detected in the image. \\
                \cmidrule(lr){1-4} 
            \textbf{Pixel Feature $C_i$} & & \\
                \cmidrule(lr){3-3} \cmidrule(lr){4-4} 
                    \phantom{aa} \multirow[t]{3}{*}{Pixel Detail Complexity} & \multirow[t]{3}{*}{$C_{i-D}$} & $_{D1}$ sharpness & A measure of the texture clarity of the image.\\
                    {} & {} & $_{D2}$ entropy & A measure of the texture randomness within the image.\\
                    {} & {} & $_{D3}$ canny\_edge & The ratio of edge pixels to overall pixels in the image.\\
                \cmidrule(lr){3-3} \cmidrule(lr){4-4} 
                    \phantom{aa} \multirow[t]{6}{*}{Pixel Aesthetic Vibrancy} & \multirow[t]{6}{*}{$C_{i-A}$} & $_{A1}$ colorfulness & An indicator of the richness and variety of colors in the image. \\
                    {} & {} & $_{A2}$ hue\_std & The standard deviation of the hue values in the image.\\
                    {} & {} & $_{A3}$ saturation\_std &  The standard deviation of the saturation values in the image. \\
                    {} & {} & $_{A4}$ lightness\_std &  The standard deviation of the lightness values in the image. \\
                    {} & {} & $_{A5}$ contrast & The intensity difference between the light and dark areas. \\
                    {} & {} & $_{A6}$ image\_variance & The variance of the pixel values in the grayscale image. \\
                \cmidrule(lr){3-3} \cmidrule(lr){4-4} 
                    \phantom{aa} \multirow[t]{3}{*}{Pixel Average Perception} & \multirow[t]{3}{*}{$C_{i-P}$} & $_{P1}$ hue & The average hue value of the pixels in the image. \\
                    {} & {} & $_{P2}$ saturation & The average saturation value of the pixels in the image. \\
                    {} & {} & $_{P3}$ lightness & The average lightness value of the pixels in the image. \\
                \cmidrule(lr){1-4}
            \textbf{Scene Recognition $S_i$} & & \\
                \cmidrule(lr){3-3} \cmidrule(lr){4-4} 
                    \phantom{aa} \multirow[t]{3}{*}{Commercial Scene} & \multirow[t]{3}{*}{$S_{i-B}$} & $_{B1}$ downtown & Probability of the image being central business district.\\
                    {} & {} & $_{B2}$ office\_building & Probability of the image being building used for business or work.\\
                \cmidrule(lr){3-3} \cmidrule(lr){4-4} 
                    \phantom{aa} \multirow[t]{3}{*}{Residential Scene} & \multirow[t]{3}{*}{$S_{i-R}$} & $_{R1}$ apartment & Probability of the image being outdoor view of apartments.\\
                    {} & {} & $_{R2}$ neighborhood & Probability of the image being area with houses and apartments.\\
                    {} & {} & $_{R3}$ food\_court & Probability of the image being dining area with multiple eateries.\\
                \cmidrule(lr){3-3} \cmidrule(lr){4-4} 
                    \phantom{aa} \multirow[t]{4}{*}{Transportation Scene} & \multirow[t]{4}{*}{$S_{i-T}$} & $_{T1}$ parking\_lot & Probability of the image being area for parking vehicles.\\
                    {} & {} & $_{T2}$ driveway & Probability of the image being private road leading to a house\\
                    {} & {} & $_{T3}$ highway & Probability of the image being main road for fast travel.\\
                \cmidrule(lr){3-3} \cmidrule(lr){4-4} 
                    \phantom{aa} \multirow[t]{4}{*}{Public Space Scene} & \multirow[t]{4}{*}{$S_{i-P}$} & $_{P1}$ plaza & Probability of the image being open public square.\\
                    {} & {} & $_{P2}$ market/outdoor & Probability of the image being outdoor market for shopping.\\
                    {} & {} & $_{P3}$ campus & Probability of the image being grounds of a university or college.\\
                    {} & {} & $_{P4}$ promenade & Probability of the image being public area for walking.\\
                \cmidrule(lr){3-3} \cmidrule(lr){4-4} 
                    \phantom{aa} \multirow[t]{4}{*}{Nature and Park Scene} & \multirow[t]{4}{*}{$S_{i-N}$} & $_{N1}$ field/wild & Probability of the image being natural open land.\\
                    {} & {} & $_{N2}$ forest\_path & Probability of the image being path through a forest.\\
                    {} & {} & $_{N3}$ forest\_leaf & Probability of the image being forest with broadleaf trees.\\
                    {} & {} & $_{N4}$ park & Probability of the image being public green area.\\
                \cmidrule(lr){3-3} \cmidrule(lr){4-4} 
                    \phantom{aa} \multirow[t]{2}{*}{Construction Scene} & \multirow[t]{2}{*}{$S_{i-C}$} & $_{C1}$ construction & Probability of the image being where buildings are being built.\\
                    {} & {} & $_{C2}$ industrial\_area & Probability of the image being industrial factory area.\\
            \bottomrule
        \end{tabularx}
    \end{table}

\newpage
\section{Rationale explanation of the survey and training sample sizes}
\label{appendix_sample_sizes}

    Given the constraints on resources and participant availability for the visual evaluation, it was essential to carefully determine an appropriate sample size. An excessively large number of images would risk incomplete evaluations, leading to inaccurate results, while too small a sample would fail to comprehensively represent the diversity of street scenes. Thus, a balanced approach was needed to ensure both feasibility and representativeness.

    Our solution was to employ stratified sampling, incorporating equal representation from five street scene categories identified through a $k$-means clustering algorithm. From each category, 100 images were randomly selected, yielding a total of 500 images. We invited 176 participants to perform 18 pairwise comparisons for each of the 19 VPI and VATA indicators. On average, each indicator was compared 6.34 times per image, resulting in 3168 pairwise comparison results per indicator. The methodology is detailed in Section~\ref{sec:method_survey}, with relevant visuals provided in Figures~\ref{method_scoring} and \ref{method_svi selection}.

\begin{figure}[!ht]
        \centering    
        \includegraphics[width=1\textwidth]{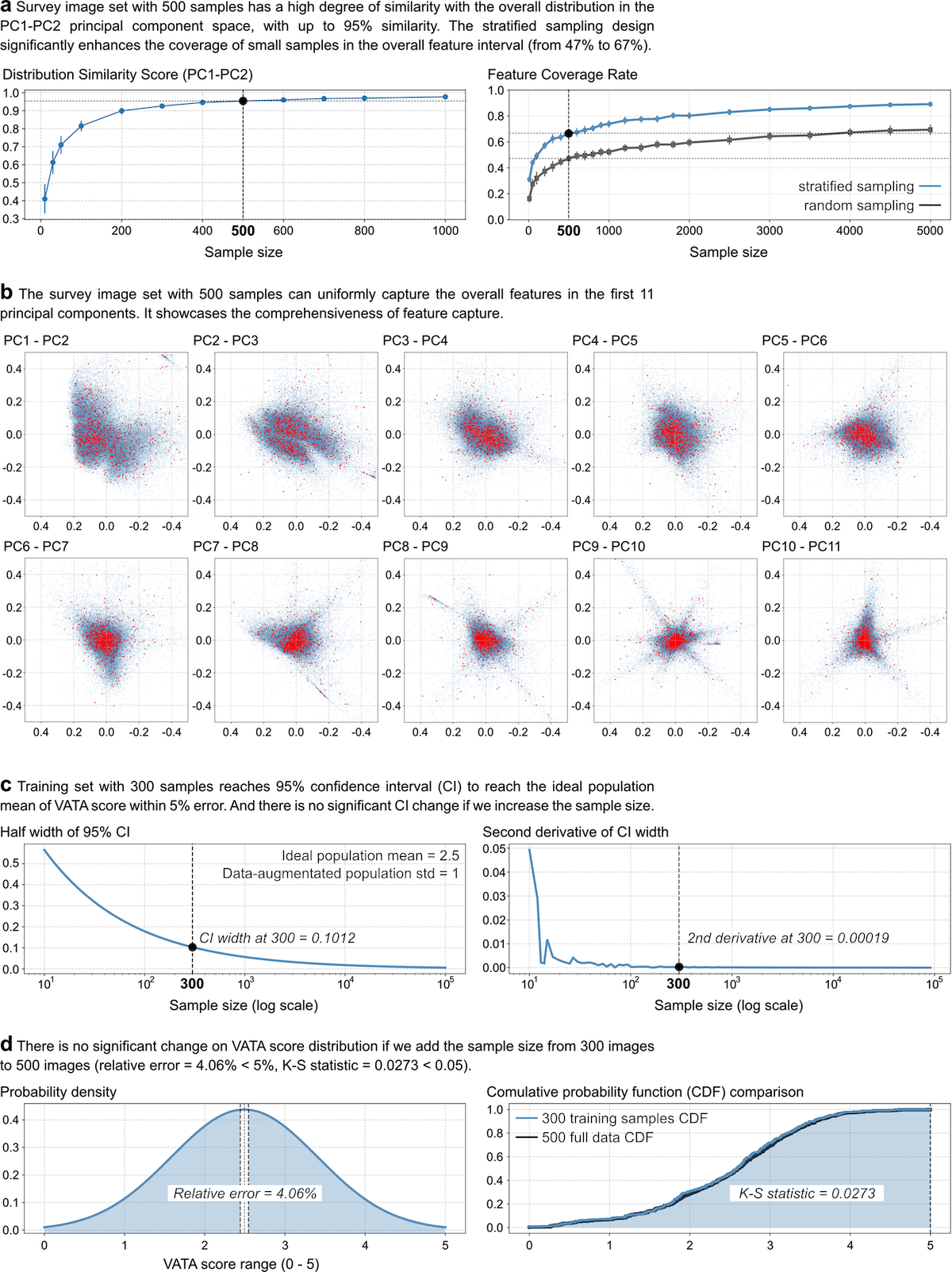} 
        \caption{Rationale analytics of survey sample size (500) and training sample size (300)}
        \label{appendix_sample size} 
    \end{figure}

    A sample size of 500 images successfully captures the diversity of street scene features, as demonstrated in Figure~\ref{appendix_sample size}. Subfigure (a) highlights that a sample of 500 images captures the overall distribution of street scenes across the first two principal components (PC1 and PC2) with a similarity of 0.95, ensuring comprehensive representation. Beyond this size, the marginal benefit of additional samples diminishes significantly. Stratified sampling further improved coverage of the feature space, increasing it from 0.47 to 0.67. This performance matches what would typically require random sampling of approximately 4,000 images. Subfigure (b) visualizes this improvement, showing that the 500 sampled images (red dots) uniformly represent the overall distribution (blue dots) across the first 11 principal components.

    Figure~\ref{appendix_sample size} Subfigures (c) and (d) demonstrate that a training sample size of 300 images is sufficient to estimate the average VATA score within a 95\% confidence interval (CI), with an error margin of only 5\%, while 200 images from the survey sample are allocated for use in the validation and test sets. This estimation assumes an ideal population mean of 2.5 (based on the normal distribution assumption of the TrueSkill algorithm) and a data-augmented population standard deviation of 1. Additionally, increasing the sample size does not result in a significant reduction in CI width, as shown by the graph illustrating the second derivative of the CI width. Moreover, as the sample size increases from 300 to 500, the distribution of VATA scores remains largely unchanged (K-S statistic $<$ 0.05), further suggesting that a representative and convergent sample of 300 images is convergent and sufficient for training a VATA prediction model.
    
\newpage
\section{Details on MTNNL model training data and process}
\label{appendix_training}

\begin{figure}[!ht]
        \centering    
        \includegraphics[width=1\textwidth]{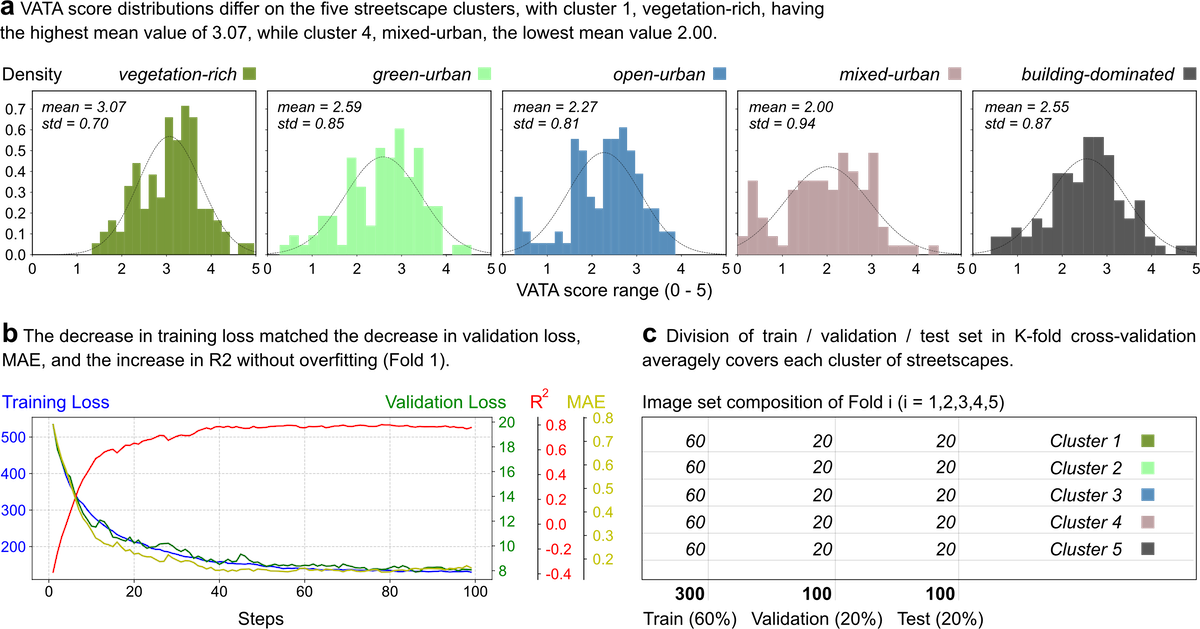} 
        \caption{Train-validation-test set division in K-folds and training process metrics.} 
        \label{method_model training} 
    \end{figure}

Figures \ref{method_model training}.a and \ref{method_model training}.c demonstrate that different streetscape categories can influence their VATA scores obtained from the online survey. During the division of train/validation/test sets, it is ensured that an equal quantity of each of the five streetscape categories is included in every fold. Figure \ref{method_model training}.b illustrates the training process for fold 1 in k-fold cross-validation model training. Both validation loss and train loss consistently decrease and stabilise, indicating that no overfitting occurred during training. 

\newpage
\section{Details on stage-1 ENRM models}
\label{appendix_enrm}
    From the stage-1 ENRM models shown in Figure \ref{appendix_stage one}, we can observe that the $R^2$ values for IF-VPI inference modelling in the microclimate inference and environmental assessment sections—except for ‘material comfort’—exceed 0.6. This suggests that these VPIs are highly predictable and can be effectively modelled using IFs. Additionally, the VPIs ‘enclosure’, ‘transparency’, and ‘complexity’ in the design quality section demonstrate strong consistency, while in the evoked emotion section, only the $R^2$ for ‘lively’ exceeds 0.6. This indicates that most VPIs related to evoked emotions vary significantly between individuals. It is crucial to consider both IF and VPI dimensions in VATA evaluation, as visual-perceptual indicators capture human visual-perceptual interpretations that are not fully reflected in SVI image features.
    \begin{table}[!ht]
        \caption{Stage-1 ENRM models for IF-VPI inference}
        \label{appendix_stage one}
        \centering
        \footnotesize
        \begin{tabularx}{\textwidth}{l>{\centering\arraybackslash}X>{\centering\arraybackslash}X>{\centering\arraybackslash}X>{\centering\arraybackslash}X}
            \toprule
            \textbf{Perception Model} & \textbf{adjusted R²} & \textbf{MAE} & \textbf{MSE} & \textbf{Significant Coefficients (out of 52 IFs)} \\
            \midrule
            \multicolumn{5}{l}{\textbf{Microclimate Inference}} \\
            \cmidrule(lr){1-1} \cmidrule(lr){2-5}
            temp\_intensity & 0.6011 & 0.4091 & 0.5279 & 24 \\
            sun\_intensity & 0.6179 & 0.3852 & 0.4887 & 24 \\
            humidity\_inference & 0.6788 & 0.2983 & 0.3571 & 38 \\
            wind\_inference & 0.6417 & 0.3557 & 0.4610 & 27 \\
            \midrule
            \multicolumn{5}{l}{\textbf{Environment Assessment}} \\
            \cmidrule(lr){1-1} \cmidrule(lr){2-5}
            traffic\_flow & 0.6571 & 0.3289 & 0.4020 & 37 \\
            greenery\_rate & 0.7994 & 0.2263 & 0.2573 & 29 \\
            shading\_area & 0.8943 & 0.1017 & 0.1123 & 43 \\
            material\_comfort & 0.4069 & 0.4576 & 0.6333 & 29 \\
            \midrule
            \multicolumn{5}{l}{\textbf{Design Quality}} \\
            \cmidrule(lr){1-1} \cmidrule(lr){2-5}
            imageability & 0.5105 & 0.4379 & 0.5660 & 37 \\
            enclosure & 0.7927 & 0.2237 & 0.2515 & 26 \\
            human\_scale & 0.5209 & 0.4446 & 0.6198 & 25 \\
            transparency & 0.6183 & 0.3843 & 0.4701 & 30 \\
            complexity & 0.7375 & 0.2805 & 0.3275 & 41 \\
            \midrule
            \multicolumn{5}{l}{\textbf{Evoked Emotion}} \\
            \cmidrule(lr){1-1} \cmidrule(lr){2-5}
            safe & 0.4235 & 0.4827 & 0.6568 & 32 \\
            lively & 0.7102 & 0.2563 & 0.3125 & 32 \\
            beautiful & 0.5598 & 0.3830 & 0.5093 & 34 \\
            wealthy & 0.4072 & 0.5116 & 0.7259 & 35 \\
            boring & 0.4378 & 0.4462 & 0.6072 & 35 \\
            depressing & 0.4733 & 0.4526 & 0.6158 & 34 \\
            \bottomrule
        \end{tabularx}
    \end{table}

\section{Feature weight analytics in the ENRM inference modelling}
\label{appendix_inference}

    \begin{figure}[!ht]
        \centering    
        \includegraphics[width=1\textwidth]{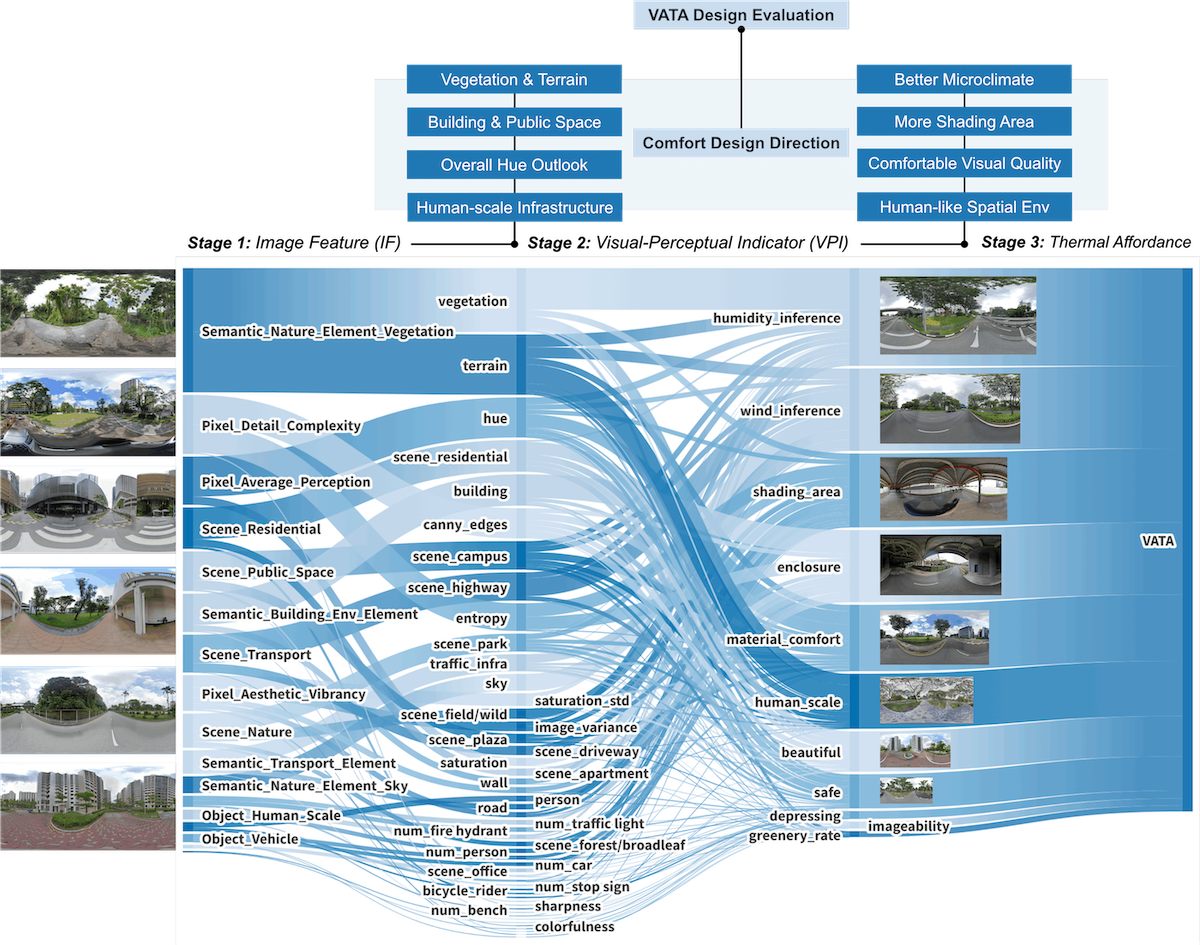} 
        \caption{Sankey diagram of positive feature weight proportion in the IF-VPI-VATA two-stage inference modelling.} 
        \label{result_VATA_positive features} 
    \end{figure}
    
    \begin{figure}[!ht]
        \centering    
        \includegraphics[width=1\textwidth]{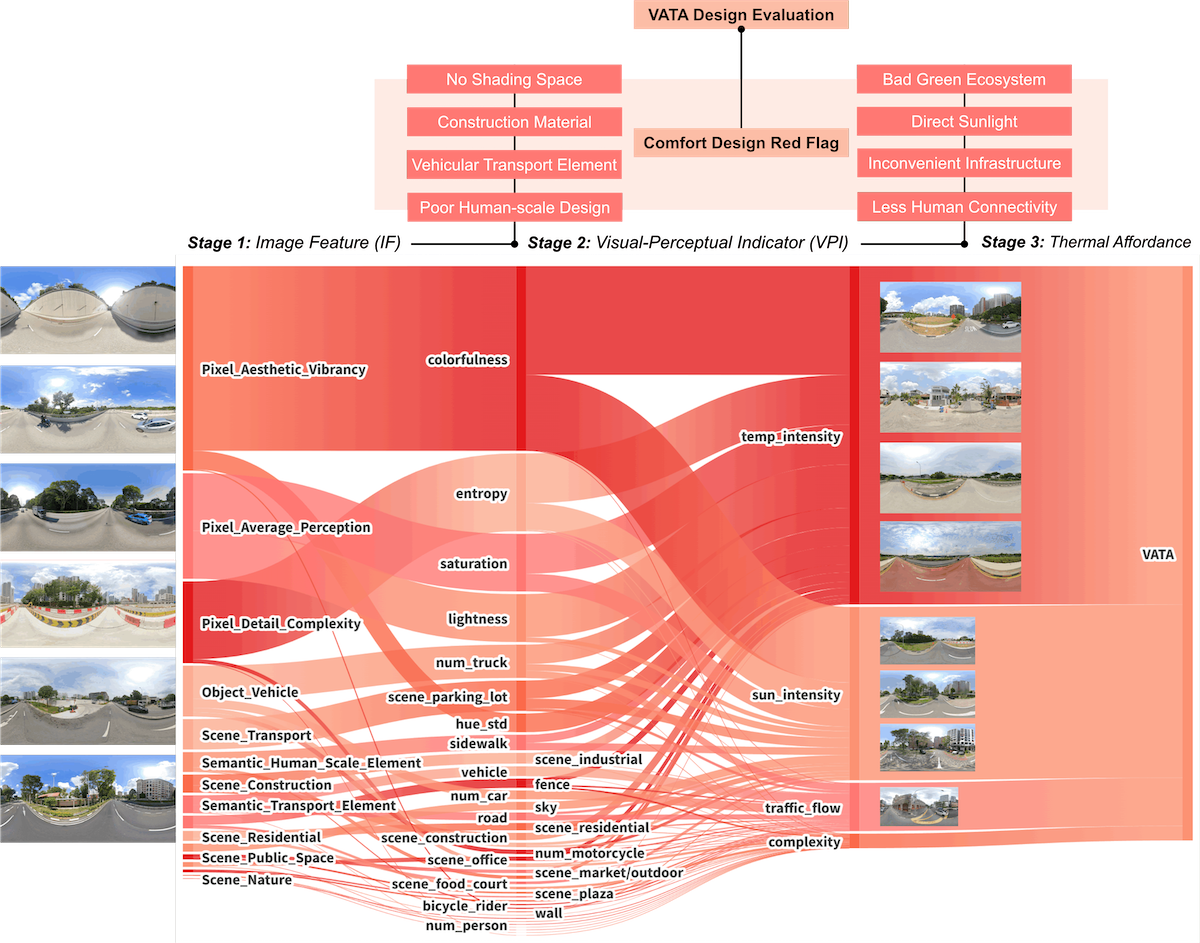} 
        \caption{Sankey diagram of negative feature weight proportion in the IF-VPI-VATA two-stage inference modelling} 
        \label{result_VATA_negative features} 
    \end{figure}
    
    The weights from the IF-VPI-VATA two-stage modelling are normalised and the proportion distributions of positive and negative weights are visualised as Sankey diagrams (Figure \ref{result_VATA_positive features} and Figure \ref{result_VATA_negative features}). Figure \ref{result_VATA_positive features} highlights the proportional distribution of the weights of positive features, where IFs and VPIs with higher weights significantly enhance the VATA level of streetscapes. Conversely, Figure \ref{result_VATA_negative features} shows the proportional distribution of the weights of negative features, where IFs and VPIs with higher absolute weights more strongly reduce the VATA level of streetscapes.

    From all positive features, semantic elements like nature (vegetation, terrain, sky), open urban scenes (residential, campus, park), and colourful pixel features (hue) dominate the IF. Key VPIs include microclimate inference (humidity, wind), environmental assessment (shading area, material comfort), design quality (enclosure, human scale), and emotional response (beautiful, safe). We summarise four important aspects of IF and VPI that are conducive to improving the thermal affordance level of streetscapes, as shown in Figure \ref{result_VATA_positive features}: 1) higher vegetation and terrain proportions can bring a better street microclimate; 2) more buildings and public spaces can increase the proportion of shading area; 3) a better overall hue outlook of streetscapes can improve visual comfort; 4) more human-scale infrastructures can make streetscape spatial environments more human-friendly and comfortable.

    From all negative features, pixel-level characteristics, traffic elements, and complex urban scenes dominate the IF, while essential VPIs include temperature intensity, sunlight intensity, traffic flow, and complexity. Four key aspects of IF and VPI that reduce thermal affordance (Figure \ref{result_VATA_negative features}) are: 1) less shading area results in more direct sunlight and thus higher temperature; 2) a higher proportion of construction materials and complicated urban scenes (transport, construction) reduce the proportion of green ecosystems in streetscapes; 3) more vehicular transport elements imply fewer pedestrian connections; 4) lack of human-scale designs indicate decreased infrastructure convenience.

    In addition to the inference logic chains hidden in the IF-VPI-VATA two-stage modelling mentioned above, more design implications can be discovered and summarised on the basis of the VATA framework in the long-term process of sustainable streetscape planning and design.

\section{Further analytics on feature-wise relationships}
\subsection{Correlation among VATA and visual-perceptual indicators (VPI)}

    \begin{figure}[!ht]
        \centering    
        \includegraphics[width=1\textwidth]{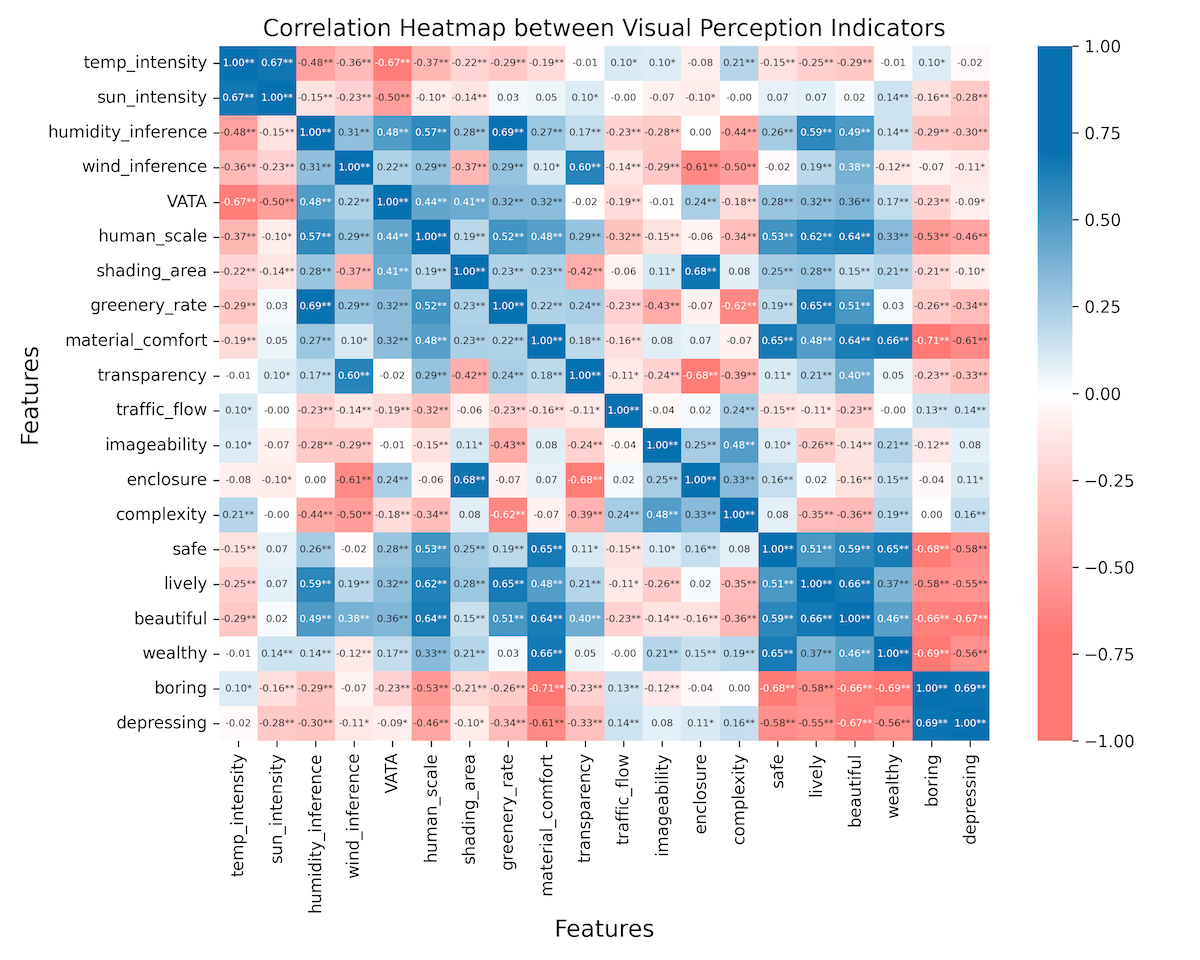} 
        \caption{Correlation coefficient among VATA and visual-perceptual indicators (VPI)} 
        \label{appendix_corr_vpi} 
    \end{figure}

Figure \ref{appendix_corr_vpi} shows the cross-correlation matrix between VATA and 19 VPIs. The results reveal strong positive relationships between VATA and several key VPIs, including humidity inference, wind inference, human scale, shading area, greenery rate, material comfort, and transparency. These correlations highlight the importance of green spaces and human-scale considerations in improving streetscape thermal affordance. Conversely, traffic flow, imageability, enclosure, and complexity exhibit negative correlations with VATA, suggesting that these elements may detract from the overall thermal affordance of urban environments.

    \begin{figure}[!ht]
        \centering    
        \includegraphics[width=1\textwidth]{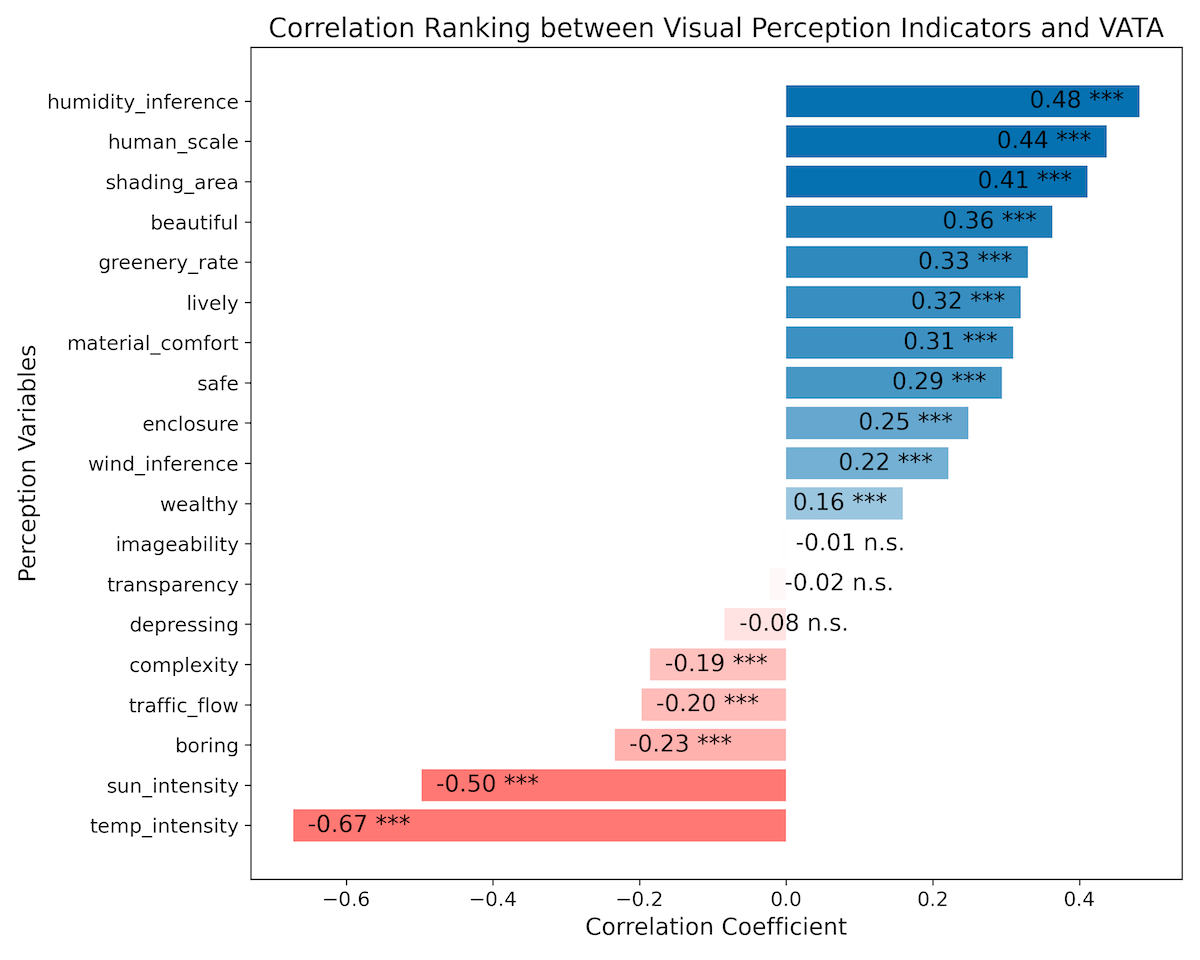} 
        \caption{Ranking of VPIs' correlation with VATS.} 
        \label{appendix_corr_vpi_vata} 
    \end{figure}

Figure \ref{appendix_corr_vpi_vata} ranks VPIs' correlations with VATA. Factors like humidity inference, human scale, shading area, 'beautiful' emotion, and greenery rate are strongly positively correlated with VATA, significantly contributing to thermal affordance. In contrast, temperature intensity, sunshine intensity, 'boring' emotion, traffic flow, and complexity are significantly negatively correlated with VATA, indicating their detrimental impact. Notably, imageability, transparency, and 'depressing' emotion show no significant correlations with VATA, suggesting that these aspects, while important in urban design, may not directly influence thermal affordance assessment. This insight invites further research into the nuanced role of urban form and aesthetics in shaping human thermal affordance perception.

\subsection{Relationship between image features (IF) and visual-perceptual indicators (VPI)}
\begin{figure}[!ht]
        \centering    
        \includegraphics[width=1\textwidth]{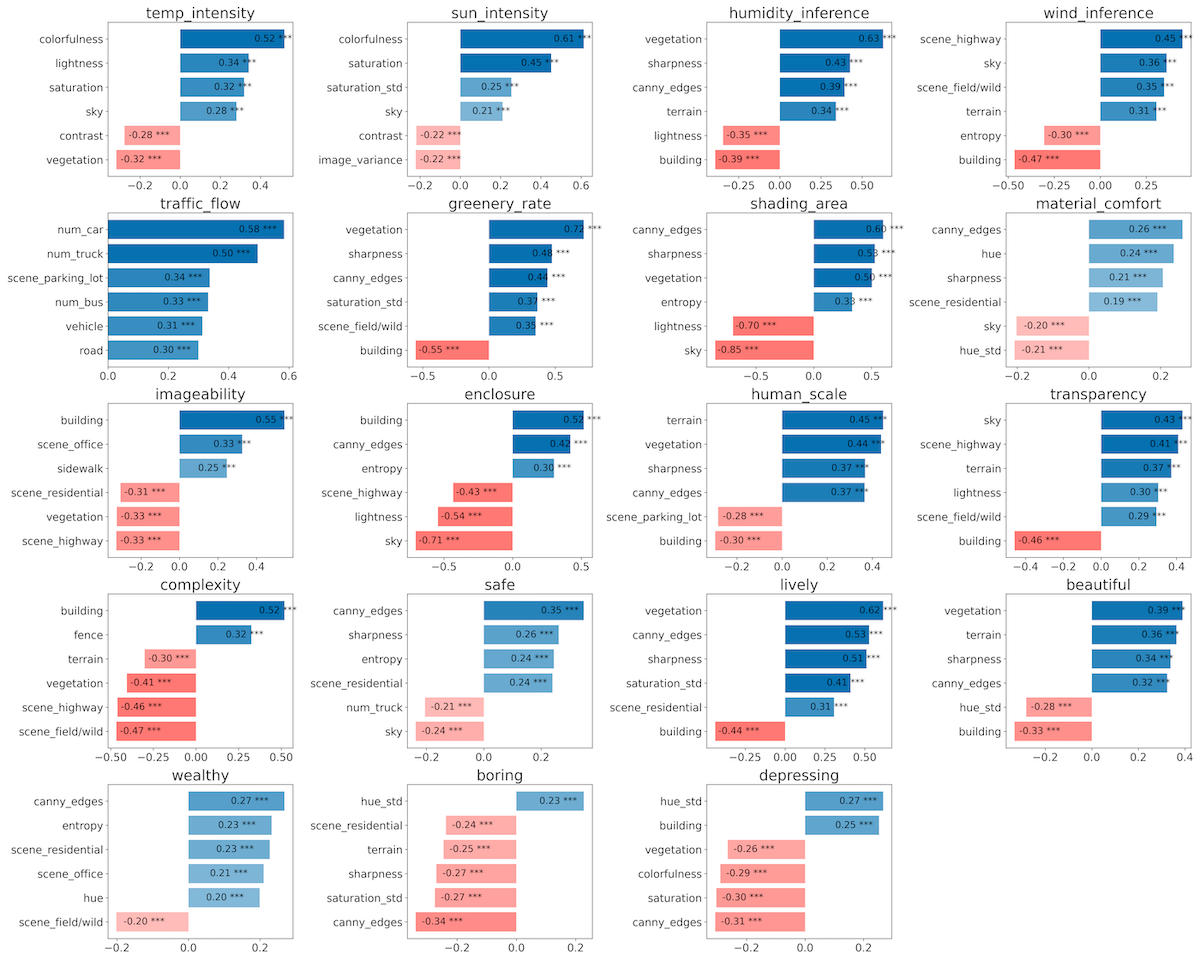} 
        \caption{Top six correlated IFs for each VPI.} 
        \label{appendix_corr_vpi_if} 
    \end{figure}

The exploration of correlations between IFs and VPIs reveals nuanced interactions that significantly influence urban thermal affordance assessment. This study incorporates a comprehensive analysis of 52 IFs and 19 VPIs, offering a detailed cross-correlation evaluation. The methodology prioritises the six most relevant IFs for each VPI providing a ranked evaluation as depicted in Figure \ref{appendix_corr_vpi_if}. 

    \begin{figure}[!ht]
        \centering    
        \includegraphics[width=1\textwidth]{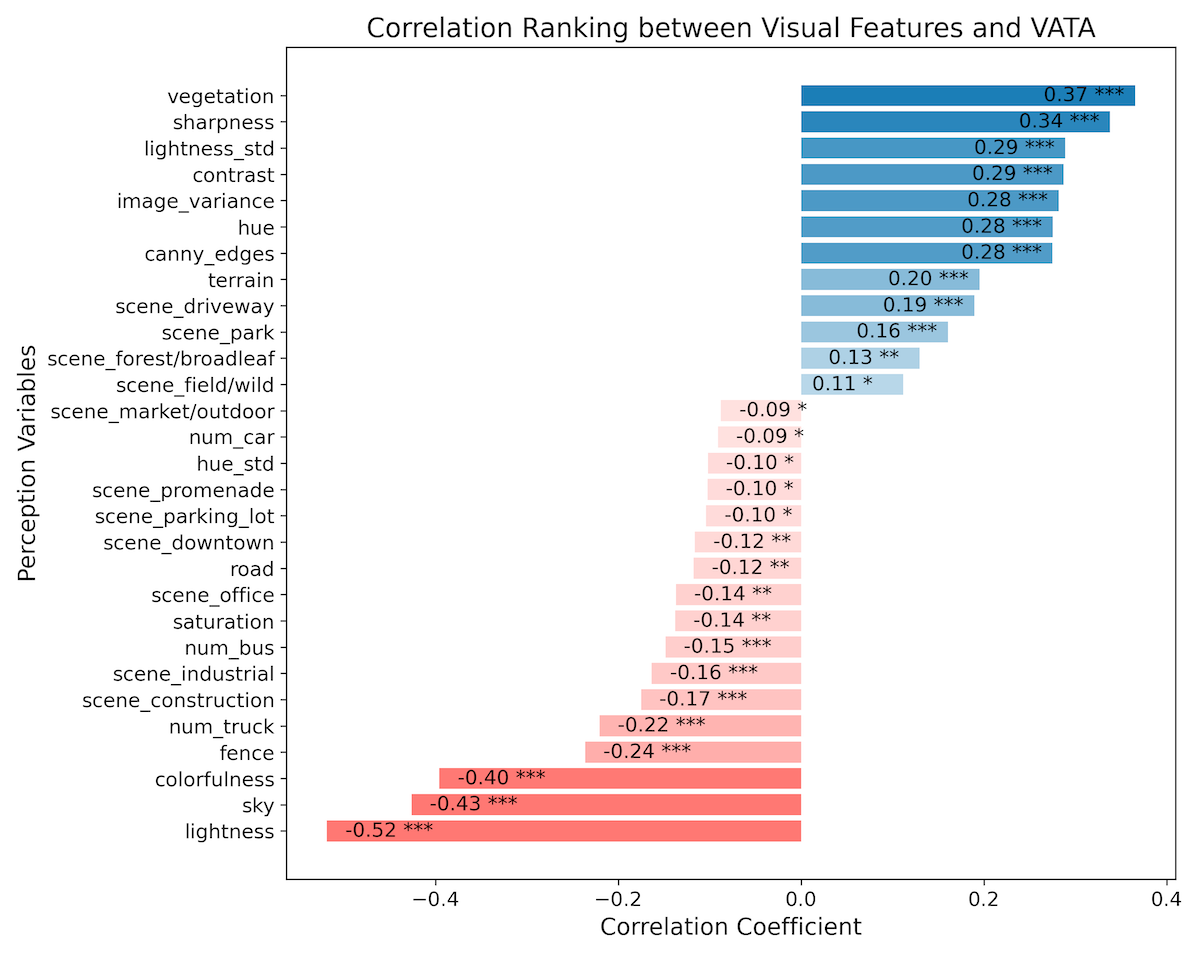} 
        \caption{Ranking of top twenty IFs' correlation with VATA.} 
        \label{appendix_corr_if_vata} 
    \end{figure}

Vegetation emerges as a pivotal element, showing strong positive correlations with humidity inference, greenery rate, shading area, and human scale. It significantly influences VATA, as shown in \ref{appendix_corr_if_vata}. However, vegetation and terrain negatively correlate with temperature intensity, traffic flow, imageability, and complexity, highlighting the complex interplay between natural and built elements in urban settings and emphasising the critical role of vegetation and terrain in shaping thermally sustainable streetscapes. Buildings present a mixed influence on VATA and VPIs. They generally negatively impact VATA evaluation, correlating positively with enclosure and complexity, and negatively with humidity inference, wind speed inference, greenery rate, and human scale. However, buildings contribute positively to the shading area, aligning with vegetation to enhance VATA.

This intricate web of correlations underscores the multifaceted nature of streetscape elements in determining visual assessment of urban thermal comfort. It also calls attention to the necessity of a balanced integration of natural and built components in urban planning to foster sustainable and comfortable living environments.

\end{document}